\documentclass[twocolumn,epjc3]{svjour3}          

\usepackage{graphicx}
\usepackage{dcolumn}
\usepackage{bm}
\usepackage{amsmath,multirow}
\usepackage{amssymb}
\usepackage{subfigure}
\usepackage{color}
\usepackage{verbatim} 
\usepackage{float}
\graphicspath{{./}{figpdf/}}

\RequirePackage[T1]{fontenc}

\smartqed  

\journalname{EPJC}

\begin{document}

\title{Third type of spacetime with the coexistence of integrability and non-integrability}


\author{
Junjie Lu  
\and Xin Wu$^{a}$} 

\thankstext{e2}{e-mail: wuxin$\_$1134@sina.com; 21200006@sues.edu.cn (corresponding author)}

\institute{School of Mathematics, Physics and Statistics, Shanghai
University of Engineering Science, Shanghai 201620, China
}

\date{Received: date / Accepted: date}

\maketitle

\begin{abstract}

The integrability or non-integrability of a spacetime usually
refers to whether the motion of massive or massless particles in
the spacetime is integrable or not. The standard black hole
spacetimes such as the Schwarzschild and Kerr metrics are always
integrable for both timelike and null geodesics. They belong to a
first type of spacetime. However, the Melvin type spacetimes as a
second type of spacetime are non-integrable, regardless of whether
they are for massive or massless particle motion. In this paper,
we discover the possibility of a third type of spacetime with
non-integrable dynamics of timelike geodesics and integrable
dynamics of null geodesics. In fact, conformal transformations may
transform type one solutions into type three. This is due to the
conformal factors preventing the separation of variables from the
Hamilton-Jacobi equation and leading to the absence of a fourth
constant of motion for the massive particle dynamics.
Nevertheless, the massless particle motion  still remains
integrable in these metrics for any conformal factors because the
conformal factors have no effect on the null geodesics whatsoever.
The conformal Kerr metric is an example of the third type of
spacetime. In addition to the conformal transformation method,
other paths may yield the third type of spacetime. The
Kerr-Bertotti-Robinson black hole metric and the accelerating
Schwarzschild spacetime are two examples of non-conformal
solutions that are also of type three.


\end{abstract}

\section{Introduction}

The motion of a massive particle or a massless particle (such as a
photon) in a spacetime is usually used to show the integrability
or non-integrability of this spacetime. If the motion is
integrable or non-integrable, this spacetime is also said to be
integrable or non-integrable.

Four standard black hole (BH) spacetimes involving the
Schwarzschild, Reissner-Nordstr\"{o}m, Kerr and Kerr-Newman
metrics are always integrable for both massive and massless
particle motion. This attributes to the existence of four
conserved quantities consisting of the massive (or massless)
particle energy, massive (or massless) particle  angular momentum,
massive (or massless) particle  rest mass and Carter constant in
each of the spacetimes. The Carter constant [1,87] arises from the
separation of variables in the Hamilton-Jacobi equation. This
constant is so important to determine not only the integrability
of massive or massless particle geodesics but also the presence of
circular and spherical massive or massless particle orbits. On one
hand, it directly brings the existence of stable equatorial and
non-equatorial circular massive particle orbits around rotating
BHs [2-5]. These stable circular massive particle orbits build up
a stable spherical massive particle orbits near each of these BHs.
There are bound massive particle orbits such as stable circular
massive particle orbits and stable spherical massive particle
orbits in these metrics. The bound orbits signify that the massive
particles neither are captured by the central bodies nor escape to
infinity; namely, they range over a finite interval of radius.
These stable orbits serve as a direct probe of the BH
gravitational fields because they are crucially dependent on some
fundamental parameters like the BH mass and spin and determine the
structure of accretion disks around the BHs. By observing matter
behavior at these orbits on the accretion disks, one can measure
these BH fundamental parameters and understand both the emission
properties of the accretion disks and the production of jets. On
the other hand, the Carter constant also induces integrable null
geodesics involving equatorial and non-equatorial circular photon
orbits (or rings) and spherical photon orbits around the BHs
[6-13]. In general, no bound photon orbits can exist in the
exterior regions of the extremal horizons of these metrics. This
fact is due to equatorial photon effective potentials having no
closed pockets but having potential barriers corresponding to
local maximum values outside these BH horizons. As a result, the
photons move according to one of three possible cases: falling to
the BHs, scattering towards infinity and doing unstable circular
orbits (such as unstable equatorial circular photon orbits) around
the BHs. These unstable circular photon orbits and spherical
photon orbits are relatively useful to analytically compute BH
shadows. By obtaining these BH shadows, one can measure the
related BH fundamental parameters and test theories of gravity
[14,15]. These spacetimes are called as a first type of spacetime
with the integrability of massive and massless particle  dynamics.

Nevertheless, there are some other BH spacetimes that are
non-integrable with respect to both massive and massless particle
motion. The fundamental reason is that the Hamilton-Jacobi
equation for each of these spacetimes loses the separation of
variables and the fourth motion constant is absent. Such a class
of typical examples are the Melvin type BH spacetimes including
Schwarzschild-Melvin, Reissner-Nordstr\"{o}m-Melvin, Kerr-Melvin
and Kerr-Newman-Melvin spacetimes [16, 17], which describe
combinations of BH gravity fields and external electromagnetic
fields. In fact, chaotic dynamics of neutral massive particles can
be found in the Schwarzschild-Melvin [18-20],
Reissner-Nordstr\"{o}m-Melvin [21], Kerr-Melvin and
Kerr-Newman-Melvin [22] spacetimes. The chaotic dynamics of
charged or neutral massive particles around some BHs are helpful
to cause acceleration and escape of ionized particles from
accretion discs and to form  relativistic jets [23]. In these
Melvin type spacetimes, null geodesics are also chaotic under some
circumstances [24]. In particular, the chaotic photon orbits can
be bound because equatorial photon effective potentials have
closed pockets. In this case, stable circular and spherical photon
orbits are also existent [25]. In addition to these Melvin type
spacetimes, two charged black holes [26], Reissner-Nordstr\"{o}m
diholes [8], rotating boson stars [27], Kerr BHs with scalar hair
[27] and Hartle-Thorne [28] spacetimes allow for the onset of
bound chaotic photon orbits. Due to the chaotic dynamics of
photons around the BHs, self-similar fractal structures appear in
BH shadows [29,30]. The non-integrability of massive and massless
particle  dynamics is also suitable for the core-shell models
[31],  black ring metric [32], Zipoy-Voorhees metric [33], Bonnor
spacetime [34], Kerr-like metric with mass quadrupole moment [35],
a system of three black holes in static equilibrium configuration
[36] and the Manko, Sanabria-G\'{o}mez, Manko metric [37]. These
spacetimes are attributed to  a second type of spacetime with
non-integrable geodesics of  massive and massless particles.

There is a third type of spacetime with the coexistence of both
non-integrability of timelike geodesics and integrability of null
geodesics. Conformal transformations to the first type of metric
with integrability of both massive and massless particle  motion
are a main path to probably obtain the third type of spacetime.
This is because null geodesics and timelike geodesics have
different demonstrations under the conformal transformations. The
former geodesics remain unchanged, whereas the latter geodesics do
not [38,39]. An example is a conformal Kerr spacetime obtained via
an invertible conformal transformation to the Kerr spacetime [40].
This conformal Kerr metric has a conformal factor with a
parity-violating interaction. The conformal factor as an effective
external force has different impacts on both massive and massless
particle  motion. It causes the timelike geodesics to be
non-integrable but does not affect null geodesics at all. In other
words, the conformal Kerr spacetime supports both the
non-integrability of timelike geodesics and the integrability of
null geodesics, as will be shown in later discussions. In this
sense, such a spacetime is viewed as the third type of spacetime.
In addition to the method of conformal transformations, other
paths may provide the third type of spacetime. Thus, the main
motivation of the present paper is to report several axisymmetric
metrics belonging to the third type of spacetime.

The rest of this paper is organized as follows. In Sect. 2, we
discuss the integrable or non-integrable dynamics of axisymmetric
metrics, which are derived from conformal transformations to the
first type of spacetime in the literature. In this way, we expect
to find the third type of spacetime allowing for the coexistence
of integrable null geodesics and non-integrable massive particle
geodesics. The conformal Kerr BH spacetime [40] as an example of
the third type of spacetime is listed in Sect. 3. As a second
example that is not a conformal metric,  a Kerr-Bertotti-Robinson
BH solution describing a Kerr BH immersed in an external uniform
magnetic field [41] is introduced in Sect. 4. An accelerating
Schwarzschild  BH metric [42], as a third example that does not
stem from a conformal transformation, is given in Sect. 5.
Finally, our conclusions are drawn in Sect. 6.

\section{A main path to obtain the third type of
spacetime}

At first, a first type of spacetime with the integrable dynamics
of massive and massless particles is introduced. Then,  a
conformal transformation is given to the first type of metric. The
equations of motion for these massive and massless particles in
the conformal metric are obtained from three different steps. In
this way, such a conformal transformation to the first type of
metric may yield a third type of spacetime with both
non-integrable massive particle dynamics and integrable null
geodesics.

\subsection{First type of spacetime}

In coordinates $x^{\mu}=(t,r,\theta,\phi)$, an axisymmetric
spacetime admitting a timelike vector field (i.e. a stationary,
axisymmetric  spacetime including a static, axisymmetric spacetime
as a special case of the  stationary, axisymmetric  spacetime) is
given by
\begin{eqnarray}
   g_{\mu\nu} &=& f(r,\theta) \bar{g}_{\mu\nu}, \\
    f(r,\theta) &=& f_1(r)+f_2(\theta),
\end{eqnarray}
where $f$ is a smooth, nonvanishing  function of $r$ and $\theta$.
$f_1(r)$ and $f_2(\theta)$ are functions of $r$ and $\theta$,
respectively.

The geodesics of a free particle with mass $m$ moving in this
gravitational field is described in terms of the Lagrangian
formalism
\begin{eqnarray}
 \mathcal{L}=\frac{m}{2}g_{\mu\nu}\dot{x}^{\mu}\dot{x}^{\nu}.
\end{eqnarray}
The 4-velocity $\dot{x}^{\mu}$ satisfies the constraint condition:
$\dot{x}_{\mu}\dot{x}^{\mu}=\varepsilon$. Here, $\varepsilon=-1$
for timelike geodesics and $\varepsilon=0$ for null geodesics.
Note that the 4-velocity normalization condition is based on the
speed of light taken as one geometric unit $c=1$. Hereafter, the
constant of gravity $G$ is set to 1, too. The Lagrangian
corresponds to generalized momenta
\begin{eqnarray}
p_{\mu}= \frac{\partial \mathcal{L}}{\partial \dot{x}^{\mu}}
=mg_{\mu\nu}\dot{x}^{\mu}.
\end{eqnarray}
Because the metric $g_{\mu\nu}$ is stationary and axisymmetric,
the generalized momenta $p_t$ and $p_{\phi}$ are constants of
motion, which correspond to the  energy $E$ and angular momentum
$L$ of massive test particle. The two motion constants are
expressed as
\begin{eqnarray}
  p_t &=& mg_{t\nu}\dot{x}^{\nu}=-E, \\
  p_\phi &=& mg_{\phi\nu}\dot{x}^{\nu}=L.
\end{eqnarray}
The momenta, energy and angular momentum are also suitable for
those of a massless particle but the mass factor should be
dropped.

With the aid of the Legendre transformation, the Lagrangian
becomes its equivalent Hamiltonian formulation
\begin{eqnarray}
H=\frac{1}{2m}g^{\mu\nu}p_{\mu}p_{\nu}=\frac{1}{2mf} \bar{H},
\end{eqnarray}
where $\bar{H}=\bar{g}^{\mu\nu}p_{\mu}p_{\nu}$. Using the
constants $E$ and $L$, the Hamiltonian system can be reduced to a
third constant of motion:
\begin{eqnarray}
H=\frac{1}{2}m\varepsilon.
\end{eqnarray}
In practice, Eq. (8) arises from the momentum constraint
$p_{\mu}p^{\mu}=\varepsilon m^2$ associated with the 4-velocity
constraint condition. Suppose that $g_{\mu\nu}$ is the first type
of metric, which is integrable for the massive particle motion.
This requires that $\bar{H}$ should have two separable parts:
\begin{eqnarray}
\bar{H} =\bar{H}_1(r,p_r)+\bar{H}_2(\theta,p_\theta).
\end{eqnarray}
Therefore, Eqs. (7)-(9) correspond to the Hamilton-Jacobi
equation, which is separable to the variables and leads to the
existence of a fourth motion constant $Q$:
\begin{eqnarray}
\varepsilon m^2f_1(r)-\bar{H}_1=\bar{H}_2-\varepsilon
m^2f_2(\theta)=Q.
\end{eqnarray}
The fourth constant is similar to the Carter constant [1,87] and
is needed to render the metric Liouville integrable [46-48]. That
is, the motion of massive or massless particle is solely regular.
Note that $m=1$ is always taken for massive particle motion in
this paper. Although $m=0$ should have been given for massless
particle motion, $m=1$ is allowed because of the introduction of
$\varepsilon=0$ or $\varepsilon=-1$.

For the Schwarzschild metric $g_{\mu\nu}$, the function $f$
consists of two separable terms
\begin{eqnarray}
f_1(r)=r^2, ~~~~ f_2(\theta)=0.
\end{eqnarray}
In this case, the equation (10) with the separability of variables
can be allowed. This leads to the existence of the constant $Q$.
Thus, the Schwarzschild metric is integrable regardless of whether
$\varepsilon=-1$ for the particle or $\varepsilon=0$ for the
massless particle. The Kerr metric $g_{\mu\nu}$ is also integrable
[1,87] due to the presence of the Carter constant $Q$ from the
separating variable equation (10) with the function $f$ split into
two pieces
\begin{eqnarray}
f_1(r)=r^2, ~~~~ f_2(\theta)=a^2\cos^2\theta.
\end{eqnarray}

As a point is illustrated, the purpose of extracting the factor
$f$ of Eq. (2) from the metric $g_{\mu\nu}$ of Eq. (1) is to find
the fourth motion constant $Q$. This metric is integrable for the
massive particle motion, and is also for the massless particle
motion. In this sense, it belongs to the first type of spacetime.

\subsection{Conformal metric}

If $\Omega$ is a smooth, strictly positive function with respect
to $r$ and $\theta$, then the metric
\begin{eqnarray}
  \tilde{g}_{\mu\nu} = F(r,\theta) g_{\mu\nu}
\end{eqnarray}
is said to be a conformal transformation [38,39] to the first type
of metric $g_{\mu\nu}$ in Eq. (1). $F(r,\theta)=\Omega^2$ is a
conformal factor, which is of course also a smooth, strictly
positive function. An important result of the textbooks [38,39] is
that null geodesics remain invariant but timelike geodesics are
generally changed under the conformal transformation.

In what follows, we give the motion equations of massive and
massless particles in the conformal metric (13) from three
different steps.

\subsubsection{Euler-Lagrange equations}

The Lagrangian of the conformal metric $\tilde{g}_{\mu\nu}$ is
\begin{eqnarray}
 \mathcal{\tilde{L}}=\frac{m}{2}\tilde{g}_{\mu\nu}\dot{x}^{\mu}\dot{x}^{\nu}.
\end{eqnarray}
The particle generalized momenta are defined as
\begin{eqnarray}
\tilde{p}_{\mu}=\frac{\partial\mathcal{\tilde{L}}}{\partial
\dot{x}^{\mu}}=mFg_{\mu\nu}\dot{x}^{\nu}.
\end{eqnarray}
As $\mathcal{L}$ does, $\mathcal{\tilde{L}}$ still has invariant
particle energy $\tilde{E}$ and angular momentum $\tilde{L}$:
\begin{eqnarray}
\tilde{E} &=& -\tilde{p}_{t}=-mFg_{t\nu}\dot{x}^{\nu}, \\
\tilde{L} &=& \tilde{p}_{\phi}=mFg_{\phi\nu}\dot{x}^{\nu}.
\end{eqnarray}

The Euler-Lagrange equations of $\mathcal{\tilde{L}}$ are
\begin{eqnarray}
 \frac{d\tilde{p}_{\mu}}{d\lambda}=\frac{\partial\mathcal{\tilde{L}}}{\partial x^{\mu}},
\end{eqnarray}
where $\lambda$ represents the proper time for the massive
particle while an affine parameter for the massless particle. Eq.
(18) is rewritten as
\begin{eqnarray}
 \ddot{x}^{\alpha}+\tilde{\Gamma}^{\alpha}_{\rho\nu}\dot{x}^{\rho}\dot{x}^{\nu}=0,
\end{eqnarray}
where $\tilde{\Gamma}^{\alpha}_{\rho\nu}$ is of the form
\begin{eqnarray}
 \tilde{\Gamma}^{\alpha}_{\rho\nu}=\Gamma^{\alpha}_{\rho\nu}
 +\frac{1}{\Omega}(\delta^{\alpha}_{\nu}\partial_{\rho}\Omega
 +\delta^{\alpha}_{\rho}\partial_{\nu}\Omega-g^{\alpha\mu}g_{\rho\nu}\partial_{\mu}\Omega).
\end{eqnarray}
Note that $\partial_{\mu}\Omega=\frac{\partial \Omega}{\partial
x^{\mu}}$. $\Gamma^{\alpha}_{\rho\nu}$ stems from the original
metric's Christoffel symbol, which represents the original
geodesic equations
\begin{eqnarray}
\ddot{x}^{\alpha}+\Gamma^{\alpha}_{\rho\nu}\frac{dx^{\rho}}
 {d\lambda}\frac{dx^{\nu}}{d\lambda}=0.
\end{eqnarray}
Eq. (19) is reexpressed as
\begin{eqnarray}
 \ddot{x}^{\alpha}+\Gamma^{\alpha}_{\rho\nu}\dot{x}^{\rho}\dot{x}^{\nu}
 =-\frac{2}{\Omega}\frac{d\Omega}{d\lambda}\dot{x}^{\alpha}+
 \frac{1}{\Omega}g^{\alpha\mu}(g_{\rho\nu}\dot{x}^{\rho}\dot{x}^{\nu})\partial_{\mu}\Omega.
\end{eqnarray}
Eq. (22) is the geodesic equations for massive and massless
particle motion in the conformal metric $\tilde{g}_{\mu\nu}$.

Because $g_{\rho\nu}\dot{x}^{\rho}\dot{x}^{\nu}=0$ for null
geodesics, Eq. (22) is
\begin{eqnarray}
 \ddot{x}^{\alpha}+\Gamma^{\alpha}_{\rho\nu}\dot{x}^{\rho}\dot{x}^{\nu}
 =-\frac{2}{\Omega}\frac{d\Omega}{d\lambda}\dot{x}^{\alpha}.
\end{eqnarray}
Through reparametrization of the form
\begin{eqnarray}
d\lambda=\Omega^2d\tilde{\lambda},
\end{eqnarray}
Eq. (23) is simplified in the following form
\begin{eqnarray}
 \ddot{x}^{\alpha}+\Gamma^{\alpha}_{\rho\nu}\frac{dx^{\rho}}
 {d\tilde{\lambda}}\frac{dx^{\nu}}{d\tilde{\lambda}}
 =0.
\end{eqnarray}
This shows that the null geodesics under the conformal
transformation coincide with those in the original metric. The
conservation of null geodesics is also supported in the two
textbooks [38,39]. Note that the parameter relation
$d\tilde{\lambda}=\Omega^2d\lambda$ used in the Wald's book is
based on the comparison of Eq. (25) with Eq. (21). In other words,
$\tilde{\lambda}$ of Eq. (25) in the Wald's work is similar to
$\lambda$ of our Eq. (21).

For timelike geodesics,
$g_{\rho\nu}\dot{x}^{\rho}\dot{x}^{\nu}=-1/\Omega^2$. This result
is because the the conformal metric is required to satisfy the
constraint $\tilde{g}_{\rho\nu}\dot{x}^{\rho}\dot{x}^{\nu}=-1$.
Equivalently, $\Omega^2g_{\rho\nu}\frac{dx^{\rho}}{d\tau}
\frac{dx^{\nu}}{d\tau}=g_{\rho\nu}\frac{dx^{\rho}}{d\tilde{\tau}}
\frac{dx^{\nu}}{d\tilde{\tau}}=-1$, where the original proper time
$\tau$ is adjusted as the conformal proper time $\tilde{\tau}$ by
\begin{eqnarray}
d\tau/\Omega\rightarrow d\tilde{\tau}\Rightarrow d\tau=\Omega
d\tilde{\tau}.
\end{eqnarray}
In terms of the adjusted proper time $\tilde{\tau}$, Eq. (22)
becomes
\begin{eqnarray}
\frac{d^2x^{\alpha}}{d\tilde{\tau}^2}+\Gamma^{\alpha}_{\rho\nu}
\frac{dx^{\rho}}{d\tilde{\tau}}\frac{dx^{\nu}}{d\tilde{\tau}}
 =-\frac{2}{\Omega}\frac{d\Omega}{d\tilde{\tau}}\frac{dx^{\alpha}}{d\tilde{\tau}}-
 \frac{g^{\alpha\mu}}{\Omega^3}\frac{\partial\Omega}{\partial x^{\mu}}.
\end{eqnarray}
No matter what reparametrization is readjusted, Eq. (27) cannot be
transformed into the form similar to Eq. (25) or (23) for the
non-constant $\Omega$. This is because  the second term in the
right-hand side of Eq. (27) is not proportional to the velocity
$\frac{dx^{\alpha}}{d\tilde{\tau}}$, and reparameterization can
only include some terms proportional to the velocity but does not
eliminate the second term. That is, conformal invariance does not
remain for timelike geodesics. The result is the same as that of
the two textbooks [38,39].

The null geodesic equations (25) and the timelike geodesic
equations (27), as the motion equations of massless and massive
particles, directly come from the Euler-Lagrange equations (25).
Our derivations of the motion equations of massive and massless
particles  clearly show the results of [38,39] that null geodesics
remain invariant under the conformal transformation, but timelike
geodesics do not.

\subsubsection{Hamiltonian canonical equations}

The Lagrangian of Eq. (14) has its equivalent Hamiltonian
formulation
\begin{eqnarray}
\tilde{H}=\frac{1}{2m}\tilde{g}^{\mu\nu}\tilde{p}_{\mu}\tilde{p}_{\nu}=\frac{1}{2mfF}
\bar{g}^{\mu\nu}\tilde{p}_{\mu}\tilde{p}_{\nu}.
\end{eqnarray}
The Hamiltonian $\tilde{H}$ has the conserved energy $\tilde{E}$
of Eq. (16) and the conserved angular momentum $\tilde{L}$ of Eq.
(17). There is still the rest mass as a third motion constant:
\begin{eqnarray}
\tilde{H}=\frac{1}{2}m\varepsilon.
\end{eqnarray}
The Hamiltonian has canonical equations
\begin{eqnarray}
\frac{dx^{\mu}}{d\tau} &=& \frac{\partial\tilde{H}}{\partial
\tilde{p}_{\mu}}=\frac{1}{m}\tilde{g}^{\mu\nu}\tilde{p}_{\nu}, \\
\frac{d\tilde{p}_{\mu}}{d\tau} &=& -
\frac{\partial\tilde{H}}{\partial
x^{\mu}}=-\frac{1}{2m}\tilde{p}_{\mu}\tilde{p}_{\nu}\frac{\partial}{\partial
x^{\mu}}(\Omega^2g_{\mu\nu}).
\end{eqnarray}
Eqs. (30) and (31) are the massive particle motion equations,
where the constraint (29) should be satisfied. They are also the
massless particle equations of motion, where the time parameter
$\tau$ is replaced with the affine parameter $\lambda$.

The canonical equations (30) and (31) are a second step to obtain
the motion equations of massive and massless particles. This shows
that both the motion equations of massive particles and those of
massless particles are the same. However, their constraints are
typically different. In the constraints (29), $\varepsilon=0$ for
the null geodesics but $\varepsilon=-1$ for the timelike
geodesics. In other words, both the null geodesics and the
timelike ones have distinct initial conditions from numerical
integrations although they have the same motion equations.

\subsubsection{Time-transformed Hamiltonian}

A  reparameterization of the time parameter does not change
physical orbits of a Hamiltonian system in phase space because the
orbits are motion curves independent of reparameterization.
However, it can change the form of the equations of motion and
break the structure of the standard Hamiltonian canonical
equations. Considering that the dynamical orbits remain unchanged
under reparameterization, many authors designed  symplectic
integrators for time-transformed Hamiltonians [43-45]. Following
this idea, we apply the method of time regularization to the
Hamiltonian (28).

Take $\tau=q_0$ as a new coordinate and its corresponding momentum
as $p_0=-\tilde{H}$. An extended phase space is given by
$(x_1,x_2,q_0; \tilde{p}_1, \tilde{p}_2, p_0)$. Applying a time
transformation function $g$ to adjust the time parameter as
\begin{eqnarray}
d\tau=g d\iota, ~~~~ g=\Omega^2,
\end{eqnarray}
we have a time transformed Hamiltonian
\begin{eqnarray}
H^{*}=g(\tilde{H}+p_0)=\frac{1}{2mf}
\bar{g}^{\mu\nu}\tilde{p}_{\mu}\tilde{p}_{\nu}-\frac{1}{2}m\varepsilon\Omega^2.
\end{eqnarray}
Because the Hamiltonian $H^{*}$ does not explicitly depend on the
new time $\iota$, it is identical to zero, that is,
\begin{eqnarray}
H^{*}=0.
\end{eqnarray}

For the null geodesics with $\varepsilon=0$, the Hamiltonian of
Eq. (33) is that of Eq. (7). The null geodesics are determined by
the canonical equations of the time-transformed Hamiltonian (33),
which are $\frac{dx^{\mu}}{d\iota} = \frac{\partial
H^{*}}{\partial
\tilde{p}_{\mu}}=\frac{1}{m}g^{\mu\nu}\tilde{p}_{\nu}$ and
$\frac{d\tilde{p}_{\mu}}{d\iota} = - \frac{\partial
H^{*}}{\partial
x^{\mu}}=-\frac{1}{2m}\tilde{p}_{\mu}\tilde{p}_{\nu}(\partial
g^{\mu\nu}/\partial x^{\mu})$ in the new time $\iota$. This shows
again invariance of null geodesics under the conformal
transformation after an adjustment of the affine parameter is
required. Because the original null geodesics are integrable, the
conformal ones are also integrable.

For the timelike geodesics with $\varepsilon=-1$, the dynamical
system of Eq. (33) is unlike that of Eq. (7) due to the influence
of the non-vanishing conformal factor $F=\Omega^2$ on the time
transformation Hamiltonian (33). In this case, the timelike
geodesics are given by the canonical equations of the
time-transformed Hamiltonian (33), which are
$\frac{dx^{\mu}}{d\iota} = \frac{\partial H^{*}}{\partial
\tilde{p}_{\mu}}=\frac{1}{m}g^{\mu\nu}\tilde{p}_{\nu}$ and
$\frac{d\tilde{p}_{\mu}}{d\iota} = - \frac{\partial
H^{*}}{\partial
x^{\mu}}=-\frac{1}{2m}\tilde{p}_{\mu}\tilde{p}_{\nu}(\partial
g^{\mu\nu}/\partial x^{\mu})-m\Omega
\frac{\partial\Omega}{\partial x^{\mu}}$ in the new time $\iota$.
Due to the existence of the term $m\Omega
\frac{\partial\Omega}{\partial x^{\mu}}$, the timelike geodesics
do not remain invariant  under the conformal transformation. In
fact, Eq. (33) with Eqs. (2), (9) and (34) becomes
\begin{eqnarray}
\varepsilon m^2F(f_1(r)+f_2(\theta))
=\bar{H}_1(r,\tilde{p}_r)+\bar{H}_2(\theta,\tilde{p}_\theta).
\end{eqnarray}
If $F=f_1-f_2>0$, then the left-hand side of Eq. (35) is
separable. In this sense, there is a fourth constant $\tilde{Q}$,
which satisfies
\begin{eqnarray}
\varepsilon m^2f^2_1(r)-\bar{H}_1(r,\tilde{p}_r)=
\bar{H}_2(\theta,\tilde{p}_\theta)+\varepsilon
m^2f^2_2(\theta)=\tilde{Q}.
\end{eqnarray}
The four constants of motion imply the integrability of timelike
geodesics in the conformal metric $\tilde{g}_{\mu\nu}$. This
result is also suitable for any one of the cases
$F=f^2_1-f_1f_2+f^2_2$,
$F=f^4_1-f^3_1f_2+f^2_1f^2_2-f_1f^3_2+f^4_2$ and
$F=(F_1(r)+F_2(\theta))/f$. Here, $F_1$ and $F_2$ are two
arbitrary smooth functions, and $F>0$ should be required.

In many other cases such as $F=f_1>0$, $F=f_2>0$ or $F=f_1f_2>0$,
the left-hand side of Eq. (35) is generally inseparable and the
fourth constant $\tilde{Q}$ is no longer existent for the massive
particle motion with $\varepsilon=-1$. Thus, the metric
$\tilde{g}_{\mu\nu}$ is non-integrable and the massive particle
motion is possibly chaotic in the metric. However, Eq. (35) is
always separable and the fourth constant $\tilde{Q}$ is existent
for the massless particle  motion with $\varepsilon=0$, no matter
what positive smooth functions are given to $F$. In this sense,
these spacetimes are said to allow the coexistence of
integrability and non-integrability. The coexistence does not mean
that the motion of a massive or massless particle  is integrable
and non-integrable together in these spacetimes, but does mean
that the motion of a massive particle is non-integrable while the
motion of a massless particle  is integrable. Such spacetimes
exist because the conformal factor $F(r,\theta)$ makes an
important contribution to the massive particle motion, whereas
becomes useless for the massless particle motion. They are called
as a third type of spacetime with the integrability and
non-integrability coexisting.

The above demonstrations provide three different steps to the
motion equations of massive and massless particles. Step 1: the
motion equations are obtained from the Euler-Lagrange equations of
the Lagrangian (14). Step 2: they are the canonical equations of
the Hamiltonian (28). Step 3: they are from the canonical
equations of the time-transformed Hamiltonian (33). The motion
equations following the three different steps should be equivalent
to describe the geodesic motions of massive and massless
particles.

\subsubsection{Remarks}

As demonstrated above, conformal transformations to the first type
of metric with integrability of both massive and massless particle
motion are a possible path to obtain the third type of spacetime.
Several marks are given as follows.

\textit{Remark 1}: \textit{Advantages of the method of time
transformation} (or regularization/reparameterization)
\textit{Hamiltonians}. The method from Step 3 is a superior
technique for finding this type of spacetime, compared with the
methods of the Euler-Lagrange equations (Step 1) and the
Hamiltonian canonical equations (Step 2). The computations of
Christoffel symbol and the derivations and comparisons of geodesic
equations under reparameterizations are cumbersome in the step 1.
The Hamiltonian canonical equations (30) and (31) between the null
geodesics and the timelike ones in the step 2 are almost
indistinguishable but have only one difference in the parameter
$\varepsilon$ of the constraint condition (29). Unlike the two
methods, the method in the step 3 has several explicit advantages.
This method is clearer to show whether null or timelike geodecics
depend on the conformal factor. This method is easier to
demonstrate whether conformal transformations preserve null and
timelike geodesics. This method is capable of providing enough
constants of motion, which sufficiently determine integrability or
non-integrability of null and timelike geodesics via Liouville's
theorem [46-48]. In contrast, Steps 1 and 2 cannot directly
achieve this task.

\textit{Remark 2}: \textit{Dramatic differences of timelike
geodesics  between reparameterizations and conformal
transformations}. Although the reparameterizations and conformal
transformations lead to changes of timelike geodesic equations,
they arise from different mechanisms. The reparameterizations must
result in changing the expressional forms of the equations of
motion but do not change the same physical orbits. The equations
of motion or the Hamiltonians can be returned to the original
forms through the inverse transformation of the time
parameterization.  However, the lack of conformal invariance for
timelike geodesic equations means changes of both the expressions
of the equations of motion (or the Hamiltonians) and the physical
orbits because the conformal factors act as extra sources which
play an essential role in the massive particle geodesics. The
conformal timelike geodesic equations are not brought back to the
original ones without the conformal factors through any time
reparameterizations. In the early literature [82-85], a long
standing debate question of whether the dynamics of the Mix-Master
Universe model is integrable or not is related to the dependence
of Lyapunov exponents on time reparametrizations. The model has
positive Lyapunov exponents corresponding to chaos for the
cosmological time [82], while zero Lyapunov exponents
corresponding to no chaos for the logarithmic time [83,84]. Motter
[85] found that chaos remains invariant under time
reparametrizations when coordinate independent chaos indicators
are used. The three-volume of the universe is not a conformal
factor but is a time transformation function. Invariance of
Lyapunov exponents is not affected by any time transformations. If
the three-volume is a conformal factor, it would have an effect on
the dynamics.

\textit{Remark 3}: \textit{Comparison between Steps 2 and 3}. The
canonical equations  are the same expressions [Eqs. (30) and (31)]
for both massive and massless particle motion in Step 2, but they
are different from the time transformed Hamiltonian (33) for the
two cases in Step 3. Do these two results conflict?  No, they do
not. The two values of $\varepsilon=0$ and  $\varepsilon=-1$ in
Eq. (29) correspond to two different types of orbits described by
the canonical equations (30) and (31). These two types of orbits
are selected with the aid of  the time transformed Hamiltonian
(33).

\textit{Remark 4}: \textit{Comparison of reparameterizations}. The
null geodesics (25) and the time transformed Hamiltonian (33) use
the same time transformation function [see Eqs. (24) and (32)].
However, the timelike geodesics (27) and the time transformed
Hamiltonian (33) adopt distinct time transformations [see Eqs.
(26) and (32)]. The time transformation (26) is based on the
massive particle 4-velocity normalization condition for the
timelike geodesics (27), but the time transformation (32) focuses
on eliminating the conformal factor $F$ in the Hamiltonian (28).
Based on specific needs, alternative time transformation functions
will be selected in the following studies.

\section{Conformal Kerr Black Hole}

In the standard model of particle physics, there are some
symmetries including the Lorentz symmetry and invariance under
charge conjugation, parity transformation and time reversal.
Parity (P) means a system's behavior through the inversion of
spatial coordinates. A parity-invariant refers to the mirror world
following the same physical laws as the original ones. However,
each individual symmetry such as the parity conservation may be
broken in a new particle sector beyond the standard model or
gravity sector. The parity-violation means physical laws without
invariance under spatial reflection  and differences in physical
processes for left-handed and right-handed configurations. By
performing an invertible conformal transformation with a
parity-violating interaction on the Kerr solution in general
relativity, Tahara et al. [40] obtained an exact rotating BH
solution  in a parity-violating gravitational theory.

In Boyer-Lindquist coordinates $(t,r,\theta,\phi)$, the Kerr BH
metric  $g_{\mu\nu}$ as a stationary, axisymmetric spacetime has
its non-vanishing components
\begin{eqnarray}
&& g_{tt}=-(1-\frac{2Mr}{\rho^2}), ~~
g_{t\phi}=-\frac{2aMr}{\rho^2}\sin^{2}\theta=g_{\phi t},  \nonumber\\
&&  g_{rr}=\frac{\rho^2}{\Delta}, ~~~~~~~~~~~~
g_{\theta\theta}=\rho^2, \\
&& g_{\phi\phi}=(r^2+a^2+\frac{2Mr}{\rho^2}a^2
\sin^{2}\theta)\sin^{2}\theta, \nonumber
\end{eqnarray}
where $M$ is the BH mass,  $\Delta=r^2-2Mr+a^2$,
$\rho^2=r^2+a^2\cos^2\theta$, and $a$ denotes the BH angular
momentum per unit mass. This metric corresponds to its
contravariant components:
\begin{eqnarray}
g^{tt} &=& \frac{g_{\phi\phi}}{g_{tt}g_{\phi\phi}-g^{2}_{t\phi}}=-
\frac{1}{\Delta\rho^2} [(r^2+a^2)^2
 - a^2\Delta\sin^2\theta],\nonumber \\
g^{rr} &=& \frac{1}{g_{rr}}=\frac{\Delta}{\rho ^2}, \nonumber \\
 g^{\theta\theta} &=&
\frac{1}{g_{\theta\theta}}=
\frac{1}{\rho ^2}, \\
g^{\phi\phi} &=& \frac{g_{tt}}{g_{tt}g_{\phi\phi}-g^{2}_{t\phi}}
=\frac{\Delta-a^2\sin^2\theta}{\Delta\rho^2\sin^2\theta}, \nonumber \\
g^{t\phi} &=& g^{\phi t}=
\frac{g_{t\phi}}{g^{2}_{t\phi}-g_{tt}g_{\phi\phi}}= -
\frac{2aMr}{\Delta\rho^2}. \nonumber
\end{eqnarray}

The authors of [40] took a  conformal factor as
\begin{eqnarray}
F= 1 + \tanh\!\left( \frac{\alpha \mathcal{P}}{\Lambda^4} \right),
\end{eqnarray}
where the Chern-Simons term $\mathcal{P}$ describes the
parity-violating conformal interaction of the form
\begin{eqnarray}
\mathcal{P} &=& - \frac{96}{\rho^{12}} M^2ar \cos\theta\,(r^2 -
3a^2 \cos^2\theta) \nonumber \\
&& \cdot (3r^2 - a^2 \cos^2\theta).
\end{eqnarray}
$\alpha$ stands for a dimensionless parameter of order unity, and
$\Lambda$ is related to some mass scale. When the factor $F$
acting on the general relativity Kerr BH solution $g_{\mu\nu}$,
the metric $\tilde{g}_{\mu\nu}$ is given in Eq. (13). The
conformal Kerr metric has its contravariant metric
\begin{eqnarray}
\tilde{g}^{\mu\nu} =\frac{1}{F} g^{\mu\nu}.
\end{eqnarray}
Because the conformal factor $F$ is always a positive function of
the Chern-Simons term, it is smooth and invertible. According to
the definition of conformal transformation in [38],  $F$ satisfies
the conformal factor condition and  the solution
$\tilde{g}_{\mu\nu}$ of Eq. (13) is a conformal transformation to
the general relativity Kerr BH solution $g_{\mu\nu}$ without
doubt.

For simplicity, the BH mass $M$ and the particle mass $m$ are
taken as geometric units: $M=1$ and $m=1$. Of course, they can
also be eliminated via scale transformations to the Hamiltonian
(28) or (33) so that this Hamiltonian becomes dimensionless. The
scale transformations are listed as follows: $r\rightarrow rM$,
$a\rightarrow aM$, $t\rightarrow tM$, $\tau\rightarrow \tau M$,
$\tilde{E}\rightarrow \tilde{E}m$, $\tilde{L}\rightarrow
\tilde{L}mM$, $\tilde{p}_r\rightarrow \tilde{p}_rm$,
$\tilde{p}_\theta\rightarrow \tilde{p}_\theta mM$,
$\alpha\rightarrow \alpha M$, $\Lambda\rightarrow \Lambda/M$,
$\tilde{H}\rightarrow \tilde{H}m$ and $\tilde{H}^*\rightarrow
\tilde{H}^*m$. In terms of these scale transformations, the
relations between the dimensionless quantities and the practical
ones are established. For instance, the dimensionless distance $r$
corresponds to the practical one $rM$, and the dimensionless
angular momentum $\tilde{L}$ is the practical one $\tilde{L}mM$.

\subsection{Regular null geodesics}

For the case of $\varepsilon=0$, an effective external force
caused by the conformal factor has no impact on null geodesics in
Eq. (33). In other words, the null geodesics in the conformal Kerr
metric $\tilde{g}_{\mu\nu}$ are completely consistent with those
in the Kerr solution $g_{\mu\nu}$ through the reparameterization.
The invariance of null geodesics is one of the conformal
transformation properties in [38,39]. The conserved quantity $Q$
of Eq. (10) with Eq. (12) as the Carter constant for massless
particles on the Kerr background shows that the null geodesics are
completely integrable and non-chaotic [1, 87].

Although both the conformal Kerr spacetime and the Kerr spacetime
have the same null geodesics and the same BH shadow boundaries,
their observed BH shadow images would be somewhat different [40].
In fact, the shadow images are also linked with the intensities of
the BH images illuminated by the accretion disks of matter fields
[29,30,49,50]. The conformal factor acts as an effective force of
external matter fields. Thus, distinct accretion disks  of matter
fields exist around the two BHs.

\subsection{Timelike geodesics}

For timelike geodesics with $\varepsilon=-1$ in the Hamiltonian
(33), Tahara et al. [40] considered equatorial circular particle
orbits in  the conformal Kerr spacetime. The conformal factor
serves as the effective external force that makes the conformal
Kerr spacetime become a large deviation from the Kerr background.
In this case, there are multiple marginally stable circular orbits
whose stability changes at some values of the radial distance $r$.
The smallest radius of the marginally stable circular orbits
corresponds to the innermost stable circular orbit. Only one
single marginally stable circular orbit (i.e. the innermost stable
circular orbit) occurs when the conformal Kerr spacetime
approaches  the Kerr background. Therefore, the equatorial
circular particle orbits are not the same in the two spacetimes.

Besides these equatorial circular particle orbits, generic
particle dynamics in the 4-dimensional phase space should have
explicit differences between the two spacetimes. The differences
lie in that the geodesics of massive particles are integrable in
the Kerr background but they are not in the conformal Kerr
spacetime. The non-integrability can be seen clearly from the
conformal factor (39) leading to inseparable form of the left-hand
side $fF$ of Eq. (35) and to the absence of the fourth motion
constant $\tilde{Q}$. This shows that the conformal factor changes
not only the form of the motion equations but also the timelike
geodesic dynamics from integrability to non-integrability. This
result also reflects the conclusion of [38,39] that conformal
transformations fail to preserve timelike geodesics. In addition
to the analytical method, a numerical integration method is
employed to prove the non-integrability of massive particle
geodesics.

\subsubsection{Construction of explicit
symplectic integrator}

For the Hamiltonian (28) or (33), the best choice of numerical
methods is a symplectic integrator [51,52], which preserves the
symplectic structure of a Hamiltonian system and shows no secular
drift in errors of the integrals of motion. An explicit symplectic
scheme is superior to an implicit one in computational efficiency.
However, the explicit symplectic scheme is generally, difficulty
available for BH metrics because of non-separation of the
variables. In recent years, our group have successfully
constructed explicit symplectic algorithms for some curved
spacetimes by multi-operator splitting and composing techniques
[53-55]. Although Hamiltonians of rotating BH metrics cannot be
directly split into several explicitly integrable terms, they can
when they are given appropriate time transformations [56,57]. In
this way, many curved spacetimes can allow for the application of
explicit symplectic integrators. Adaptive time step explicit
symplectic methods have also been developed in [20]. In what
follows, we design an adaptive time step explicit symplectic
integrator for the present Hamiltonian.

For the conformal Kerr metric, the Hamiltonian (28) becomes of the
form
\begin{eqnarray}
\tilde{H} = W(r,\theta)
+\frac{1}{2}\tilde{g}^{rr}\tilde{p}^{2}_{r}
+\frac{1}{2}\tilde{g}^{\theta\theta}\tilde{p}^{2}_{\theta},\label{eq:Hamiltonian}
\end{eqnarray}
where function $W(r,\theta)$ reads
\begin{eqnarray}
W(r,\theta) =\frac{1}{2}(\tilde{g}^{tt}\tilde{E}^2
+\tilde{g}^{\phi\phi}\tilde{L}^2)
-\tilde{g}^{t\phi}\tilde{E}\tilde{L}.
\end{eqnarray}

The above Hamiltonian is not split into several explicitly
integrable terms. However, it is when an appropriate time
transformation  is implemented according to the ideas of Refs.
[56,57]. When the time transformation (32) is replaced by
\begin{eqnarray}
d\tau=g_1(r,\theta)d\tau_1,~~~~~g_1(r,\theta)=F \frac{{{\rho
^2}}}{\Delta},
\end{eqnarray}
the Hamiltonian (33) is modified as
\begin{eqnarray}
K &=& g_1(r,\theta)(\tilde{H}+p_0) \nonumber \\
&=&
g_1(r,\theta)(W(r,\theta)+p_0)+\frac{1}{2}g_1(r,\theta)\tilde{g}^{rr}\tilde{p}^{2}_{r} \nonumber \\
&&
+\frac{1}{2}g_1(r,\theta)\tilde{g}^{\theta\theta}\tilde{p}^{2}_{\theta},
\end{eqnarray}
The time transformation Hamiltonian like the Hamiltonians in
[53-55] is split into three explicitly integrable terms:
\begin{eqnarray}
K =K_1+K_2+K_3,
\end{eqnarray}
where the three parts are expressed as
\begin{eqnarray}
K_1 &=& g_1(r,\theta)(W(r,\theta)+p_0), \nonumber \\
{K_2} &=& \frac{1}{{2\Delta }}p_\theta ^2, \nonumber \\
{K_3} &=& \frac{1}{2}p_r^2. \nonumber
\end{eqnarray}

Let symplectic operators $\psi_i$ be analytical solvers of the
sub-Hamiltonians $K_i$. A first-order approximation to the exact
solution of the Hamiltonian $K$ is
\begin{eqnarray}
\chi(h) =\psi_3(h)\times\psi_2(h)\times\psi_1(h),
\end{eqnarray}
and its adjoint reads
\begin{eqnarray}
\chi^*(h) =\psi_1(h)\times\psi_2(h)\times\psi_3(h),
\end{eqnarray}
where $h$ is a time step of the new time $\tau_1$. The two
operators can compose a second-order explicit symplectic method
\begin{eqnarray}
S2(h)=\chi(\frac{h}{2})\times\chi^*(\frac{h}{2}).
\end{eqnarray}
Although the time transformation (44) is used, the two times
$\tau$ and $\tau_1$ are approximately equal when $r$ is
sufficiently large because $g_1\approx 1$ for
$r\rightarrow\infty$. Therefore, the integrator S2 does not use
adaptive time steps.

Following the idea of Ref. [20], we readjust the Hamiltonian (45)
to construct our desired method through a time reparameterization.
This readjusted Hamiltonian is of the form
\begin{eqnarray}
\Gamma &=&
\frac{K}{\Phi}+g_1\ln\left(\frac{\Phi}{\varphi(\tau)}\right)
\nonumber \\
&=& \frac{1}{\Phi}(K_1+K_2+K_3)+\Gamma_1+\Gamma_2.
\end{eqnarray}
Here, $\Phi$ rather than $p_0=1/2$ is a generalized momentum of
the time coordinate $\tau$, $\varphi=j/r$ ($j$ being a free
parameter) is viewed as a function of the original time $\tau$,
$\Gamma_1=g_1\ln\Phi$ and $\Gamma_2=-g_1\ln\varphi$. In fact, a
second time transformation is adopted via $\tau_1\rightarrow
\tau_2$: $d\tau_1=g_2d\tau_2=(1/\Phi)d\tau_2$. Thus, the time
transformation with function $g_2=rg_1/j$ from $\tau$ to $\tau_2$
is written as
\begin{eqnarray}
d\tau=g_1d\tau_1=\frac{g_1}{\Phi}d\tau_2=\frac{r}{j}g_1d\tau_2=g_2d\tau_2.
\end{eqnarray}
Taking $\Psi_1$ and $\Psi_2$ as solvers of the sub-Hamiltonian
terms $\Gamma_1$ and $\Gamma_2$, we have another explicit second
order symplectic scheme
\begin{eqnarray}
AS2(h) &=& \Psi_1(\frac{h}{2})\times \Psi_2(\frac{h}{2})\times
S2(\frac{h}{\Phi}) \nonumber
\\ && \times\Psi_2(\frac{h}{2})\times\Psi_1(\frac{h}{2}).
\end{eqnarray}
As shown in Eq. (51), the original time step $\Delta\tau$
decreases with the radial distance $r$ decreasing when the new
time step $\Delta\tau_2$ remains unchanged. This shows the
application of adaptive time steps to the method AS2.

The adaptive method AS2 and the nonadaptive one S2 have no secular
drifts in Hamiltonian errors $\Delta H =\tilde{H}+1/2$  in Fig. 1.
This is a manifestation of the advantages of symplectic
integrators. In addition, the errors of AS2 reach 8 to 12 orders
of magnitude (note that an $n$ order means $10^{-n}$), and those
of S2 are 4 to 8 orders of magnitude. That is,  the accuracy of
AS2 is about four orders of magnitude higher than that of S2.
Algorithm AS2 yields such a better enough accuracy that it can
provide reliable numerical results. Thus, AS2 is selected in the
later numerical computations.

\subsubsection{Chaotic massive particle dynamics}

For the parameter $\alpha=70.5$, Fig. 2(a) shows that several
orbits are chaotic because the points of these orbits are randomly
spread to some areas on Poincar\'{e} sections. There are also
several chains of islands. Each chain of islands is a regular
orbit consisting of three islands. In each chain of three islands,
successive points jump from the first island to the second one and
from the second one to the third one. They return from the third
one to the first one and continue to jump. These multi-island
chains will lead to the occurrence of resonance. Connection of the
three islands, i.e. occurrence of resonance overlap, forms an
orange orbit with three saddle points. Instability of the saddle
points makes the orange orbit chaotic. These results show that
regular orbits and chaotic ones are allowed together for the same
parameters. However, chaos dies out and all orbits are almost
quasi-periodic Kolmogorov-Arnold-Moser (KAM) tori when the
parameter  decreases to $\alpha=15.35$ in Fig. 2(b). Here, the
three-island chain orbits are still present. These facts seem to
show that a decrease of the parameter $\alpha$ leads to weakening
the chaoticity of massive particle orbits. This result is easily
known from the second term of Eq. (33). In fact, the term is an
increasing function of $\alpha$. The increase of $\alpha$ causes
that of the effective external force and enhances the extent of
chaos. Inversely, the decrease of $\alpha$ suppresses the extent
of chaos.

In addition to the method of Poincar\'{e} sections, the largest
Lyapunov exponent that measures the average exponential deviation
of two nearby orbits is a common qualitative tool to distinguish
between regularity and chaoticity of orbits. The variational
method and the two-particle one are two algorithms for the
calculation of Lyapunov exponent [58]. The latter method is
simpler in the application of general relativity. If a bounded
orbit has a positive Lyapunov exponent, it is chaotic; whereas it
is regular when the largest Lyapunov exponent is equal to (or
smaller than) zero.  Based on this point, $\alpha=70.5$ yields
chaotic dynamics but $\alpha=15.35$ corresponds to regular
dynamics in Fig. 2(c).

Chaos is a strong indication of the non-integrability of timelike
geodesics in the conformal Kerr spacetime. It can be concluded
from the null geodesics and the timelike geodesics that the
conformal Kerr spacetime is not integrable for the massive
particle dynamics but it is for the massless particle dynamics.
Without doubt, the conformal Kerr metric is the third type of
spacetime.

\section{Kerr-Bertotti-Robinson  metric}

The theoretical analysis in Sect. 2 and the numerical test in
Sect. 3 have well shown that conformal transformations to the
first type of spacetime provide a possible path to the emergence
of the third type of spacetime.  Of course, there are also other
paths to yield the third type of spacetime. For example, such a
metric for the Weyl tensor of algebraic type D with a non-aligned
(and non-null) electromagnetic field without sources, i.e. a
Kerr-Bertotti-Robinson (KBR)  BH solution, can be obtained from
the Einstein-Maxwell equations [41].

In fact, the KBR solution as a stationary, axisymmetric spacetime
is the Kerr BH immersed in an external uniform magnetic field
oriented along the BH rotation axis. It is described in terms of
Eq. (13), where $F$ is of the function
\begin{eqnarray}
F &=& 1/[1+B^2r^2 -B^2\Delta_1 \cos^2\theta],\\
\Delta_1  &=& \left(1-B^2M^2\frac{I_2}{I^2_1}\right)r^2-
2M\frac{I_2}{I_1}r +a^2, \nonumber \\
I_1 &=& 1-\frac{1}{2}a^2B^2, \nonumber \\
I_2 &=& 1-a^2B^2, \nonumber
\end{eqnarray}
and $g_{\mu\nu}$ has nonzero components:
\begin{eqnarray}
g_{tt}&=& \frac{1}{\rho^2}\left(-N+a^2P\sin^2\theta \right),\\
g_{rr}& =& \frac{\rho ^2}{N},\\
g_{\theta\theta}& =& \frac{\rho^2}{P},\\
g_{\phi\phi}&= &\frac{1}{\rho^2}\left[P(r^2+a^2)^2\sin^2\theta-Na^2\sin^4\theta \right],\\
g_{t\phi}&=& g_{\phi t} = \frac{1}{\rho^2} a\sin^2\theta[N -
P(r^2+a^2)].
\end{eqnarray}
Note that $P=1+B^2(M^2 I_2/I^2_1-a^2)\cos^2\theta$, $N=(1+
B^2r^2)\Delta_1$ and $B$ is the strength of external uniform
magnetic field. The contravariant components of $g_{\mu\nu}$ are
\begin{eqnarray}
g^{tt} &=& - \frac{1}{PN\rho^2}[P(r^2+a^2)^2-Na^2\sin^2\theta],\\
g^{t\phi}&=&  \frac{a}{PN\rho^2}[N-P(r^2+a^2)]=g^{\phi t},\\
g^{\phi\phi}&=&  \frac{1}{PN\rho ^2\sin^2\theta}(N-a^2P\sin^2\theta),\\
g^{rr}&=& \frac{N}{\rho^2},\\
g^{\theta \theta}&=& \frac{P}{\rho^2}.
\end{eqnarray}
The magnetized KBR BH metric $\tilde{g}_{\mu\nu}$ can also be
found in [49-51]. This metric contains three parameters consisting
of the BH mass $M$ and rotation $a$, and the magnetic field
strength $B$. If $B=0$, then $F=1$, and
$\tilde{g}_{\mu\nu}=g_{\mu\nu}$ is the usual rotating Kerr black
hole. For $B\neq0$, $F$ is not a conformal factor of the KBR
metric because the KBR metric does not stem from a conformal
transformation to the Kerr metric. However, the transformation
$g_{\mu\nu}\rightarrow\tilde{g}_{\mu\nu}$ would be conformal if
$F$ is a positive, smooth function and $g_{\mu\nu}$ is a metric of
a certain BH (excluding the Kerr BH). In this sense, $F$ seems to
be a conformal factor.

The KBR spacetime combined the gravity field and electromagnetic
field  is asymptotically finite and uniform. It has bounded
ergoregions, and can allow massive particles to escape to
infinity. On the contrary, the magnetic fields in Melvin type
spacetimes [16,17] decrease as the distances increase and
ergoregions remain unbounded, while massive particle geodesics do
not reach infinity. These properties seem to show that these
Melvin type spacetimes are not relatively realistic models to
represent the combined fields between BH gravity fields and
external electromagnetic fields. Therefore, the KBR spacetime
appears as a promising alternative to the Melvin-type of
solutions.

When a scale transformation is given to the magnetic field
$B\rightarrow B/M$ and the other quantities still employ the same
scale transformations mentioned above, the Hamiltonian (33) with
Eqs. (53)-(58) becomes dimensionless. Here, the function $f$ is
still the sum of the two terms of Eq. (12), and $\bar{g}_{\mu\nu}$
obtained from Eq. (1) is of the expression
$\bar{g}_{\mu\nu}=g_{\mu\nu}/f$.

\subsection{Massless particle  dynamics}

For massless particle  dynamics of $\varepsilon=0$, the left-hand
side of Eq. (35) is zero while the right-hand side has been
separable. This means the presence of the fourth constant
$\tilde{Q}$ in Eq. (36). As a result, the massless particle
dynamics in the KBR BH spacetime is integrable. This is the reason
why the analytical method can be used to provide an insight into
the null geodesics and BH shadows in the KBR metric [59,60].

Although the function $F$ of Eq. (53) becomes useless in the null
geodesics, the magnetic field $B$ appears in both $F$ and
$g_{\mu\nu}$ unlike the parameters $\alpha$ and $\Lambda$ that
exist only in the conformal factor $F$ of Eq. (39). Consequently,
the magnetic field greatly affects the null geodesics. In fact,
the shadow size increases with the magnetic field increasing
[59,60].

Unlike those in the KBR metric, the massless particle dynamics are
non-integrable in the Schwarzschild-Melvin,
Reissner-Nordstr\"{o}m-Melvin, Kerr-Melvin and Kerr-Newman-Melvin
spacetimes [20,24,25]. Besides unstable photon orbits like the
falling orbits, the escaping orbits and unstable circular orbits,
bound photon orbits such as bound photon circular, quasiperiodic,
chaotic orbits were found in the Kerr-Newman-Melvin spacetime
[24]. Such bound chaotic photon orbits also exist in some other
spcetimes [8,26-28]. The chaotic motion of photons gives rise to
self-similar fractal structures in BH shadows [25,29,30].

\subsection{Massive particle dynamics}

For the motions of massive particles with $\varepsilon=-1$, Eqs.
(12) and (53) clearly show the non-separation of the term $fF$ in
Eq. (35). The non-separation of timelike geodesics rather than
null geodesics was also mentioned in [86]. Therefore, no fourth
constant can be provided and the massive particle dynamics are not
integrable in the KBR spacetime. The non-integrability of massive
particle dynamics can be confirmed through the numerical method.

The adopted integrator is still AS2 of Eq. (52), where the time
transformation function of Eq. (44) should be modified as
\begin{eqnarray}
g_1(r,\theta)=\frac{F\rho^2}{PN}.
\end{eqnarray}
Accuracy of the Hamiltonian $\tilde{H}$ in the present problem is
still similar to that in Fig. 1.

Fig. 3 displays the onset of chaos of massive particles in the KBR
spacetime under some circumstances involving appropriate values of
the magnetic field $B$. There are also regular KAM tori including
chains of three islands. As $B$ decreases, chaos gets from strong
to weak and more regular orbits increase. This is because the
decrease of the magnetic field leads to weakening the
gravitational force. A notable point is that the four values of
$B$ in Fig. 3 are much smaller than $B=0.261$ as the constrained
value of the magnetic field for $a=0.8003$ in Table 1 of [59]. The
magnetic fields of the KBR BH are constrained in terms of the
images of the supermassive BHs M87* and Sgr A* from EHT
observations. The chaotic massive particle dynamics fully show the
non-integrability of the KBR spacetime.

Without question, the Melvin type BH spacetimes [16,17] allow for
the chaotic dynamics of massive particles. It was reported that
chaos of massive particles occurs in these spacetimes [18-22].
Note that the external magnetic field $B$ is so strong to change
the geometries of both the KBR BH metric and the Melvin type BH
spacetimes. Intuitively,  $B$ is inset into these metrics and
makes an important contribution to the chaoticity of massive
particles. If the external magnetic field is too small to affect
any spacetime geometry, it would play an important role in the
motion of charged massive particles and even causes the charged
massive particle dynamics to be chaotic (see e.g. [61-73]). Of
course, the dynamics of massive particles in other metrics can
exhibit chaotic behavior [31-37].

In short, the above demonstrations sufficiently show that the KBR
BH metric as the third type of spacetime allows for both the
non-integrability of massive particle dynamics and the
integrability of massless particle  dynamics. However, neither the
dynamics of massive particles nor the dynamics of massless
particles are integrable in the second type of spacetime, such as
the Melvin type BH spacetimes [16-22,24], Reissner-Nordstr\"{o}m
diholes metric [8], rotating boson stars metric [27] and
Hartle-Thorne spacetime [28].

\section{Accelerating Schwarzschild  black hole}

The C-metric [42] that does not come from a conformal
transformation to the Schwarzschild  metric is another example for
obtaining the third type of spacetime.

It  describes a class of accelerating black holes. The
acceleration represented by a parameter $A$ means that two black
holes accelerate away from each other. In spherical-type
coordinates, Eq. (13) for the description of an accelerating and
non-rotating black hole (i.e. accelerating Schwarzschild black
hole) corresponds to the following forms
\begin{eqnarray}
F &=& 1/(1+Ar\cos \theta)^2, \\
g_{tt} &=& -(1-A^2r^2)\left(1-\frac{2M}{r}\right)=-Q_1, \\
g_{rr} &=& \frac{1}{Q_1}, \\
g_{\theta\theta} &=& r^2/(1+2MA\cos\theta)=r^2/P_1, \\
g_{\phi\phi} &=& P_1r^2\sin^2\theta.
\end{eqnarray}
The metric $\tilde{g}_{\mu\nu}$ corresponds to contravariant
components
\begin{eqnarray}
\tilde{g}^{tt} &=& \frac{1}{F}\frac{1}{g_{tt}}, \quad\quad
\tilde{g}^{rr} = \frac{1}{F}\frac{1}{g_{rr}}, \nonumber \\
\tilde{g}^{\theta\theta} &=&
\frac{1}{F}\frac{1}{g_{\theta\theta}}, \quad\quad
\tilde{g}^{\phi\phi} = \frac{1}{F}\frac{1}{g_{\phi\phi}}.
\nonumber
\end{eqnarray}
The accelerating Schwarzschild metric $\tilde{g}_{\mu\nu}$ being a
static, spherical-symmetric spacetime as a special case of
stationary, axisymmetric spacetimes has two horizons at $r=2M$ and
$r=1/A$. When $A=0$, $g_{\mu\nu}$ denotes the Schwarzschild
metric.

As claimed above, the accelerating Schwarzschild metric is not
from a conformal transformation to the Schwarzschild metric
because $g_{\mu\nu}$ includes the non-vanishing acceleration
parameter $A$  and is not the Schwarzschild metric. However,
$g_{\mu\nu}\rightarrow\tilde{g}_{\mu\nu}$ \textit{would} be a
conformal transformation and the function $F$ \textit{would} be
viewed as a conformal factor if $1+Ar\cos \theta>0$ and
$g_{\mu\nu}$ is thought as a certain metric. In this sense,
function $F$ would be said to be a non-conformal factor.

The above-mentioned dimensionless operations are also suitable for
the present problem, where $A\rightarrow A/M$.

\subsection{Regular motion of photons}

For photons as one type of massless particle with $\varepsilon=0$,
the function $f$ of Eq. (2) is $f=r^2$ in Eq. (11). The fourth
integral $\tilde{Q}$ in Eq. (28) with Eq. (29) exists and is
rewritten as
\begin{eqnarray}
\frac{r^2 \tilde{E}^2}{Q_1}-r^2 Q_1\tilde{p}^2_r &=& \tilde{Q}, \\
P_1\tilde{p}^2_\theta+\frac{\tilde{L}^2}{P_1\sin^2\theta}&=&
\tilde{Q}.
\end{eqnarray}
This implies the integrability of massless particle motions. Note
that the constant $\tilde{Q}$ arises from the separation of the
Hamilton-Jacobi equation. In fact, $\tilde{p}_r=dS_r/dr$ and
$\tilde{p}_\theta=dS_\theta/d\theta$, where $S_r$ and $S_\theta$
are two terms of the generating function
\begin{eqnarray}
S =-\tilde{E}t+\tilde{L}\phi+S_r(r)+S_\theta(\theta).
\end{eqnarray}
Thus, Eqs. (70) and (71) become
\begin{eqnarray}
&& \frac{dS_{r}}{dr}=\pm\frac{\sqrt{R(r)}}{rQ_1},
\\
&&
\frac{dS_{\theta}}{d\theta}=\pm\frac{\sqrt{\Theta(\theta)}}{P_1},
\end{eqnarray}
where $R(r)$ and $\Theta(\theta)$ are expressed as
\begin{eqnarray}
R(r) &=& r^2\tilde{E}^2 -\tilde{Q}Q_1,
\\
\Theta(\theta) &=& \tilde{Q}P_1 -
\frac{\tilde{L}^2}{\sin^2\theta}.
\end{eqnarray}

Two impact parameters are defined by
\begin{eqnarray}
\xi =\frac{\tilde{L}}{\tilde{E}},  ~~~~~~
\eta=\frac{\tilde{Q}}{\tilde{E}^2}.
\end{eqnarray}
The conditions for a photon ring (or circular orbit) at any plane
$\theta\in (0,\pi)$ are
\begin{eqnarray}
R(r)&=& 0, ~~~~ \frac{dR}{dr}= 0, \\
\Theta(\theta) &=& 0.
\end{eqnarray}
The conditions for a photon sphere are Eq. (78) and
\begin{eqnarray}
\Theta(\theta)\geq0,~~{\rm i.e.,} ~~ \xi \geq\sqrt{\eta
P_1}\sin\theta.
\end{eqnarray}
Based on Eq. (78), the photon ring has its radius
\begin{eqnarray}
r_{c}=\frac{\sqrt{1+12A^2}-1}{2A^2}.
\end{eqnarray}
It is clear that $r_{c}$ increases as $A$ decreases. Of course,
$r_{c}$ is also the radius of the photon spherical orbit. When
$A\rightarrow 0$, $r_{c}=3$ is the radius of the Schwarzschild
photon ring. Using Eqs. (78) and (81), we have a critical impact
parameter
\begin{eqnarray}
\eta_c = \frac{r_{c}^2}{Q_1(r_{c})}.
\end{eqnarray}
In light of Eq. (79), another impact parameter at the plane
$\theta$ is
\begin{eqnarray}
\xi_{\theta}=\sqrt{\eta_c P_1(\theta)}\sin\theta.
\end{eqnarray}
It is also one special case of Eq. (80). Particular for  the
equatorial photon ring ($\theta=\pi/2$), another critical impact
parameter is
\begin{eqnarray}
\xi_c =\xi_{\frac{\pi}{2}}=\sqrt{\eta_c
P_1(\frac{\pi}{2})}=\sqrt{\eta_c}
=\frac{r_{c}}{\sqrt{Q_1(r_{c})}}.
\end{eqnarray}
It is a decreasing function of $A$.

For an arbitrary value of $\theta$, the impact parameter
$\xi_{\theta}$ of Eq. (83) is always solved from Eq. (79). This
means that each value of $\xi_{\theta}$ corresponds to a photon
ring at the plane $\theta$. When $\theta$ runs from 0 to $\pi$,
all these photon rings collectively form the photon sphere.
$\xi_c$ is the maximum one of all these values of $\xi_{\theta}$.
In this way, there are countless pairs of $(\xi_{\theta},\eta_c)$
(corresponding to the photon sphere). By projecting these pairs
into an observer's screen with the distance $r_o$ and angle
$\theta_o$, we can obtain innumerable points in the observer's
screen. These points form the boundary of a BH shadow. The BH
shadow is a circle with the radius $r_{\text{sh}}=\xi_c$ in the
accelerating spacetime. The radius depends on the parameter $A$
and increases with a decreases of $A$, but is independent of the
observation angle. In the Schwarzschild limit ($A=0$), the shadow
radius reduces to $r_{sh}=\sqrt{27}\approx 5.196$. Under the
assumption of Gaussian uncertainties with $-0.125 \lesssim \delta
\lesssim 0.005 ~(1\sigma)$,  a constraint on the Schwarzschild BH
shadow radius from Sgr A*'s shadow is given in [74,75] by
\begin{eqnarray}
4.55  \leq r_{\rm sh} \leq 5.22.
\end{eqnarray}
Considering this, we estimate the upper bound of the acceleration
parameter
\begin{eqnarray}
|A| \lesssim 0.0319.
\end{eqnarray}

Null geodesics in the accelerating rotating type BH spacetimes are
also integrable in general.  Signatures of shadows of accelerating
Kerr BH [76], accelerating Kerr-Newman BH [77] and accelerating
Kerr BH with a cosmological constant [78] are investigated.

\subsection{Motion of massive particles}

Because of the function $F$ of Eq. (65) and $\varepsilon=-1$, the
term $-fF$ in the left-hand side of Eq. (35) is explicitly
inseparable and the Hamiltonian $\tilde{H}$ (28) with Eqs. (13),
(65)-(69) is non-integrable.

Choosing the time transformation function (44) as
\begin{eqnarray}
g_1(r,\theta)=\frac{F}{P_1Q_1},
\end{eqnarray}
we still apply  AS2 of Eq. (52) to solve the Hamiltonian (50).
Numerical results show the presence of chaotic massive particle
orbits in Fig. 4(a). Here the accelerating parameter $A=0.000111$
is obtained in the constrained interval of Eq. (86). There are
also regular KAM tori including chains of two islands. Chaos
becomes weak and more regular orbits increase when the
accelerating parameter decreases in Fig. 4 (a)-(d). A decrease of
the accelerating parameter gives rise to an increase of repulsive
force and then weakens degree of chaos. The existence of chaos
powerfully supports the non-integrability of timelike geodesics in
this spacetime.

The accelerating Kerr BH spacetime is also non-integrable for the
motion of massive particles. In fact, chaotic massive particle
dynamics was shown in [79]. In addition, other accelerating BH
spacetimes such as the Pleba\'{n}ski-Demia\'{n}sky metric [80] and
accelerating Kerr-Newman-AdS black holes [81]  will be believed to
be likely chaotic for massive particle motions.

In a word, the dynamics of massive particles are not integrable
but the dynamics of massless particles are in many accelerating BH
spacetimes.

\section{Summary }

Conformal transformations to the first type of spacetime with the
integrability of both massive and massless particle motions are an
importantly possible path to obtain the third type of spacetime
with non-integrable massive particle dynamics and integrable
massless particle  dynamics. In fact, the choice of conformal
factors determines whether these conformal metrics are integrable
or not. Under some values of the conformal factors, these
spacetimes still remain integrable for both the motion of massive
test particles and the motion of massless particles. On the other
hand, the massive particle motions are likely non-integrable in
these spacetimes for some other values of the conformal factors.
This is due to the conformal factors preventing the separation of
variables from the Hamilton-Jacobi equation and leading to the
absence of a fourth constant of motion. However, the massless
particle  motions  are still integrable in these metrics for any
values of the conformal factors  because the conformal factors do
not affect the massless particle motion whatsoever. In this sense,
these spacetimes as the third type of spacetime allow for the
coexistence of both non-integrable massive particle dynamics and
integrable massless particle dynamics. In essence, the emergence
of the third type of spacetime is due to the basic properties of
conformal transformations that allow for invariance of null
geodesics rather than timelike geodesics.

The conformal Kerr BH metric is a typical example of the
coexisting spacetimes. In fact, it becomes the Kerr BH metric for
the massless particle  motion. The existence of the Carter
constant as the fourth motion constant results in integrable null
geodesics in the Kerr spacetime. The massless particle motions
outside the horizons include falling orbits, escaping orbits and
unstable circular orbits. The fourth constant provides a chance to
analytically calculate the boundary of Kerr BH shadows. The BH
spin exerts an important influence on the shape of the Kerr BH
shadows. However, the conformal factor serves as the effective
external force, which causes the massive particle dynamics in the
conformal Kerr spacetime and those in the Kerr background to have
great differences. A typical difference between them is that the
massive particle dynamics are not integrable in the conformal Kerr
spacetime but are in the Kerr background. This non-integrability
of timelike geodesics is numerically shown by finding chaos of
timelike geodesics in the conformal Kerr spacetime.  A decrease of
one of the parameters in the conformal factor weakens the
chaoticity of massive particle orbits.

In addition to the conformal transformation method, other paths
may induce the third type of spacetime. An example is the
magnetized KBR BH metric, which is exactly derived from the
Einstein-Maxwell equations. There are three parameters involving
the BH mass and rotation, and the external magnetic field
strength. When the magnetic field vanishes, this metric becomes
the usual rotating Kerr black hole. For the nonzero magnetic
field, the KBR metric with the factor $F$ of Eq. (53) does not
arise from a conformal transformation to the Kerr metric. The
magnetic field has a much greater impact on null geodesics because
the metric components of Eqs. (54)-(58) contain the magnetic field
strength $B$ although the factor $F$ depending on the magnetic
field parameter $B$  fails to affect the null geodesics. In spite
of this, such a magnetic field does not destroy the integrability
of the metric without the factor $F$. The KBR BH shadow size
increases as the magnetic field increases. However, the chaoticity
of massive particles is strengthened with the magnetic field
increasing. This sufficiently supports the non-integrability of
massive particle dynamics in the KBR BH metric. Unlike the KBR
metric as the third type of spacetime, the Melvin type spacetimes
as the second type of spacetime allow for the non-integrability of
both massive and massless particle dynamics.

Another example that does not stem from a conformal transformation
is the accelerating Schwarzschild BH metric. In this spacetime,
the null geodesics are integrable and the BH shadow gets smaller
when the accelerating parameter becomes larger. However, the
chaoticity of massive particles is strengthened. In other words,
the accelerating Schwarzschild BH metric is also one of the third
type of spacetime.

\textbf{Acknowledgements}: The authors are very grateful to a
referee for valuable comments and suggestions. This research is
supported by the National Natural Science Foundation of China
(Grant No. 12573077).

\textbf{Data Availability Statement} This manuscript has no
associated data or the data will not be deposited. [Author's
comment: The datasets generated and/or analysed during the current
study are available from the corresponding author on reasonable
request.].

\textbf{Code Availability Statement} Code/software will be made
available on reasonable request. [Author's comment: The
code/software generated during and/or analysed during the current
study is available from the corresponding author on reasonable
request.].

\textbf{Conflicts of Interest}: The authors declare no conflict of
interest.

\begin{figure*}[htpb]
\centering{
\includegraphics[width=20pc]{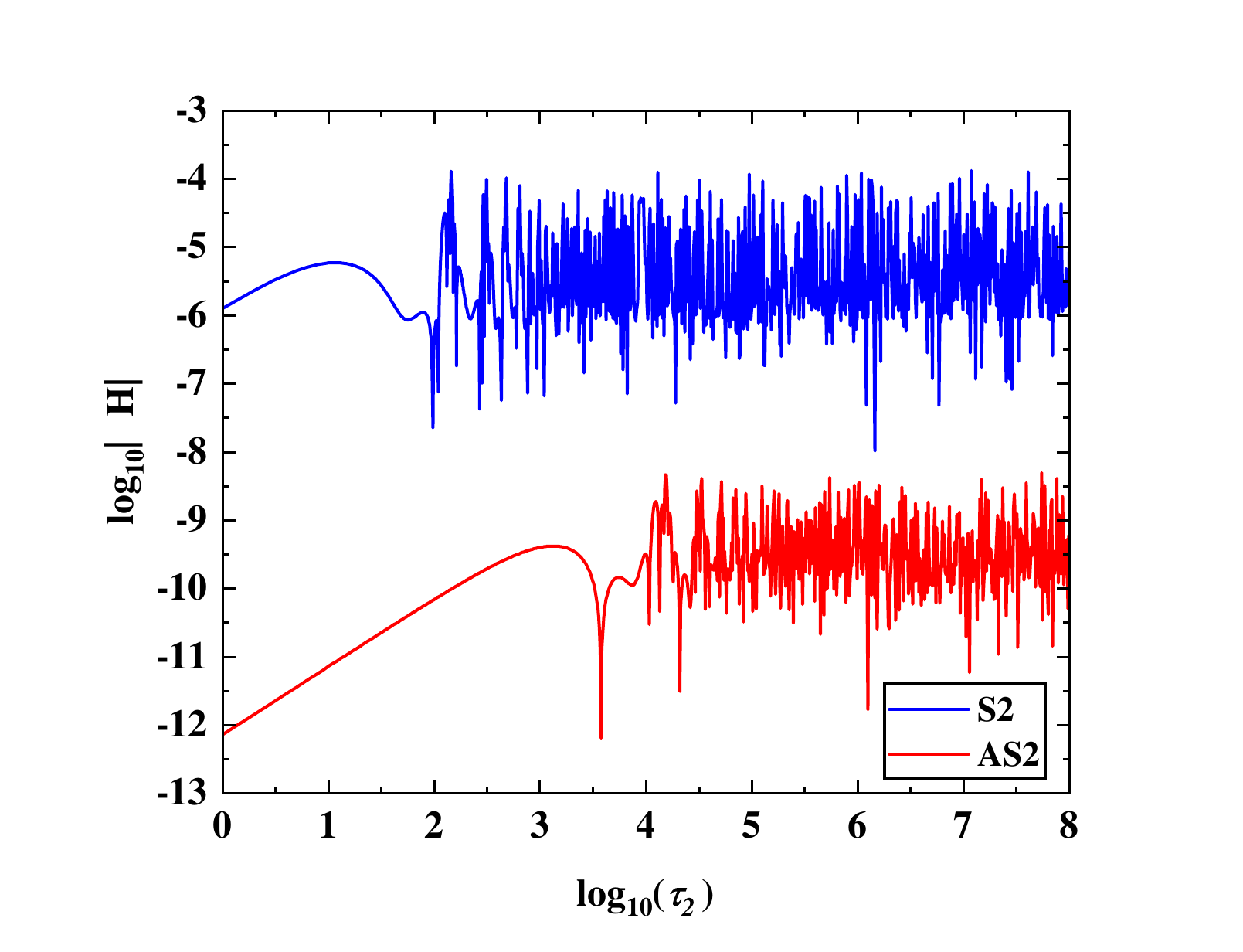}
\caption{Hamiltonian errors $\Delta H =\tilde{H}+1/2$ for the
adaptive time step explicit symplectic method AS2 and the
nonadaptive one S2. A fixed time step in the new time $\tau_2$ is
$h=1$, and $j=1000$ is adopted. The parameters are given by
$a=0.5$, $\tilde{E}=0.95$, $\tilde{L}=3.1$ and $\alpha = 70.5$. A
massive particle orbit in the conformal Kerr metric has its
initial conditions $r=8.1$, $\theta=\pi/2$ and $\tilde{p}_r=0.1$.
The initial value $\tilde{p}_\theta
>0$ is solved from Eq. (33) with Eq. (34). } }
\end{figure*}

\begin{figure*}[htpb]
    \centering{
\includegraphics[width=13pc]{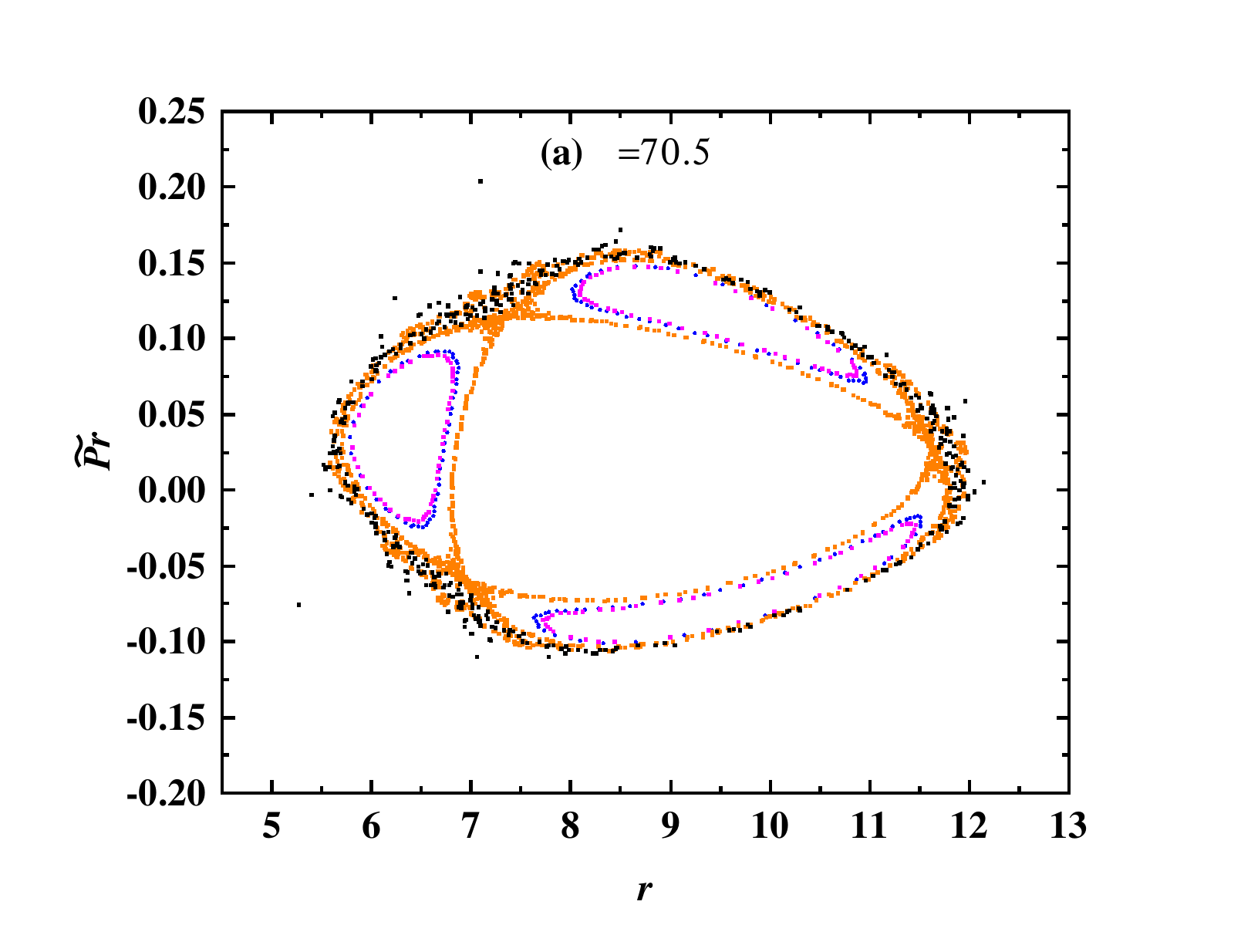}
\includegraphics[width=13pc]{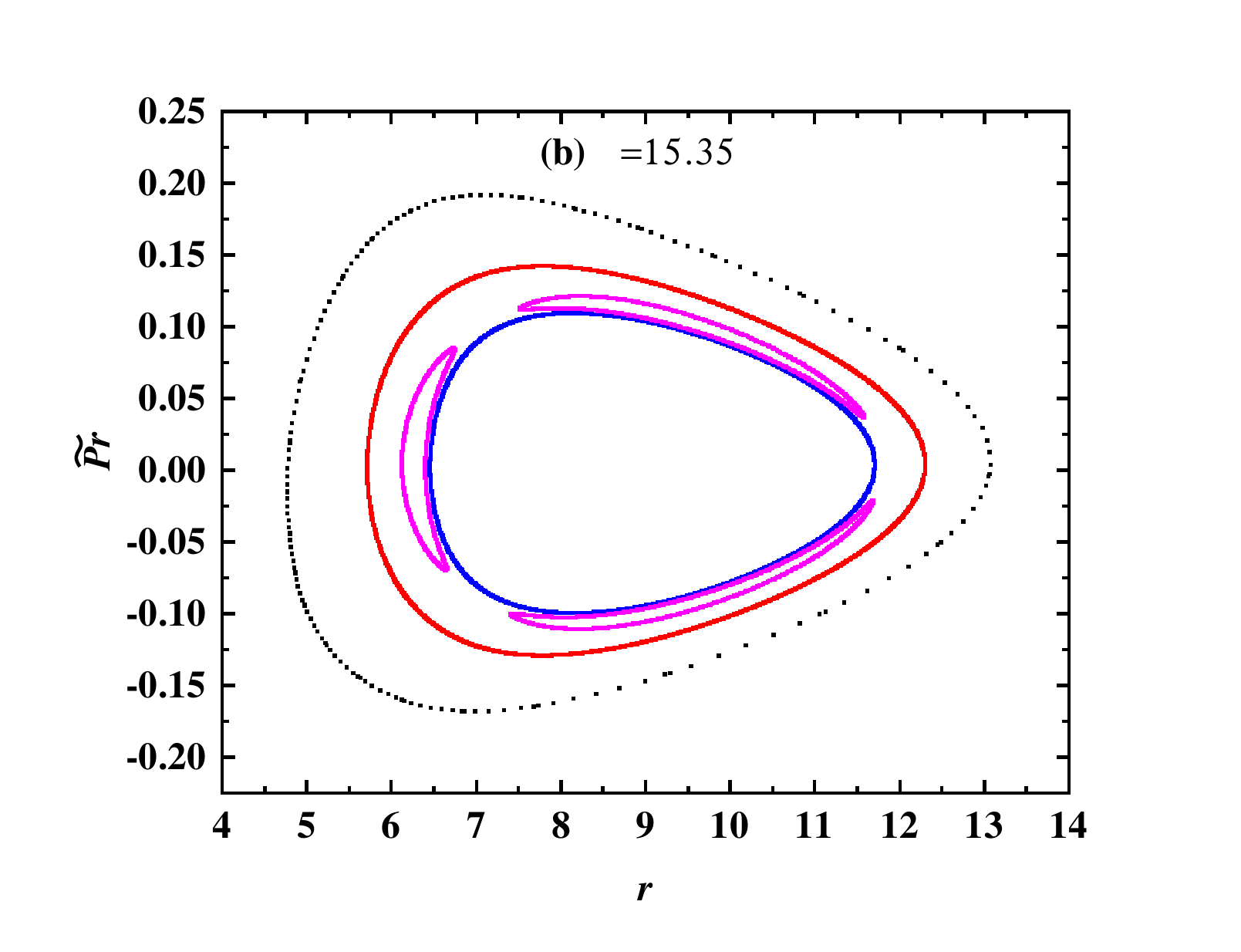}
\includegraphics[width=13pc]{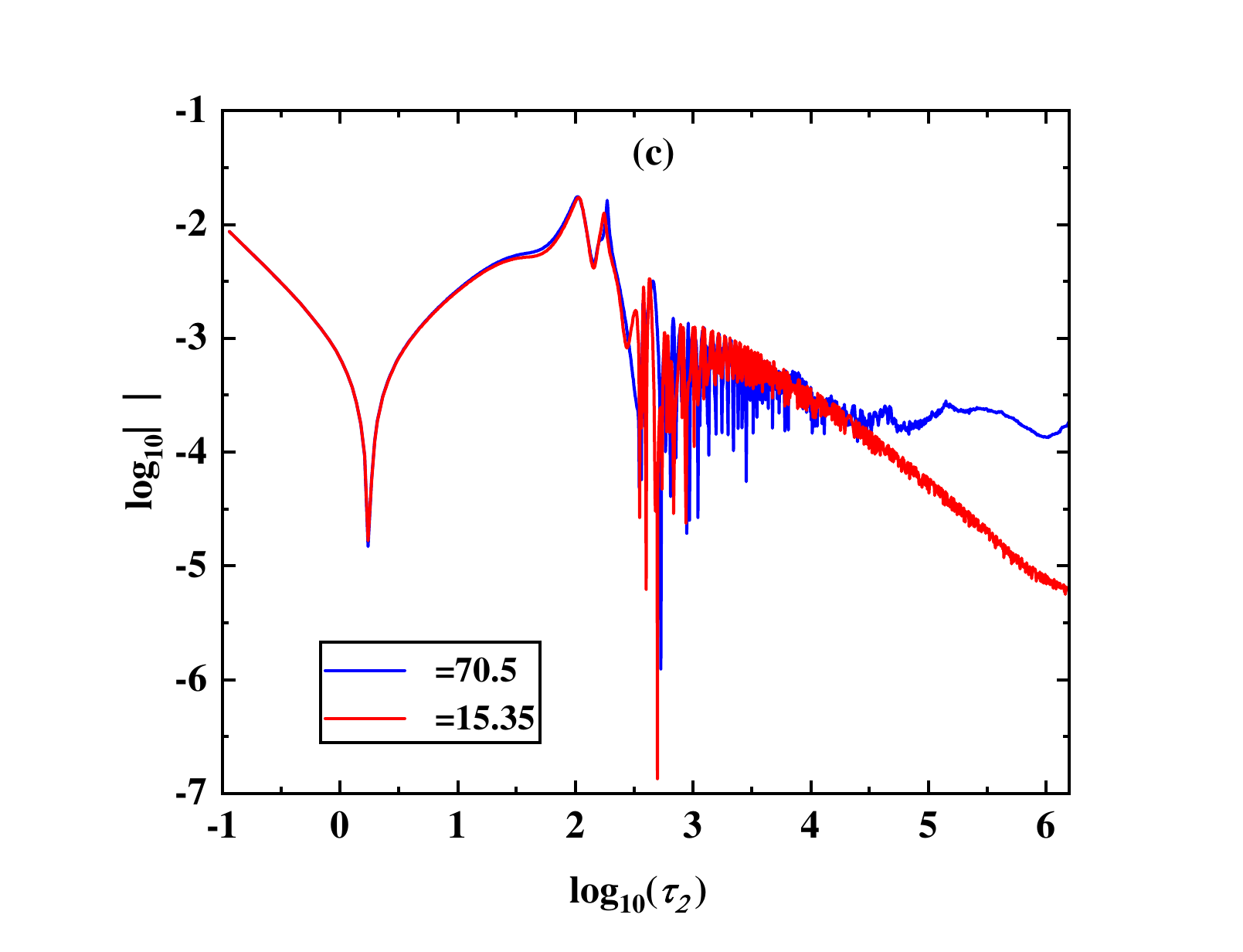}
\caption{ Poincar\'{e} sections at the plane $\theta=\pi/2$ with
$p_\theta>0$. The parameters in panel (a) are the same as those in
Fig. 1, but $\alpha=15.35$ is used in panel (b). (c) Lyapunov
exponents $\gamma$ for two values of the parameter $\alpha$. The
initial separation is $r=8.94$. The other initial conditions and
the other parameters are those of Fig. 1.} }
\end{figure*}

\begin{figure*}[htpb]
    \centering{
    \includegraphics[width=20pc]{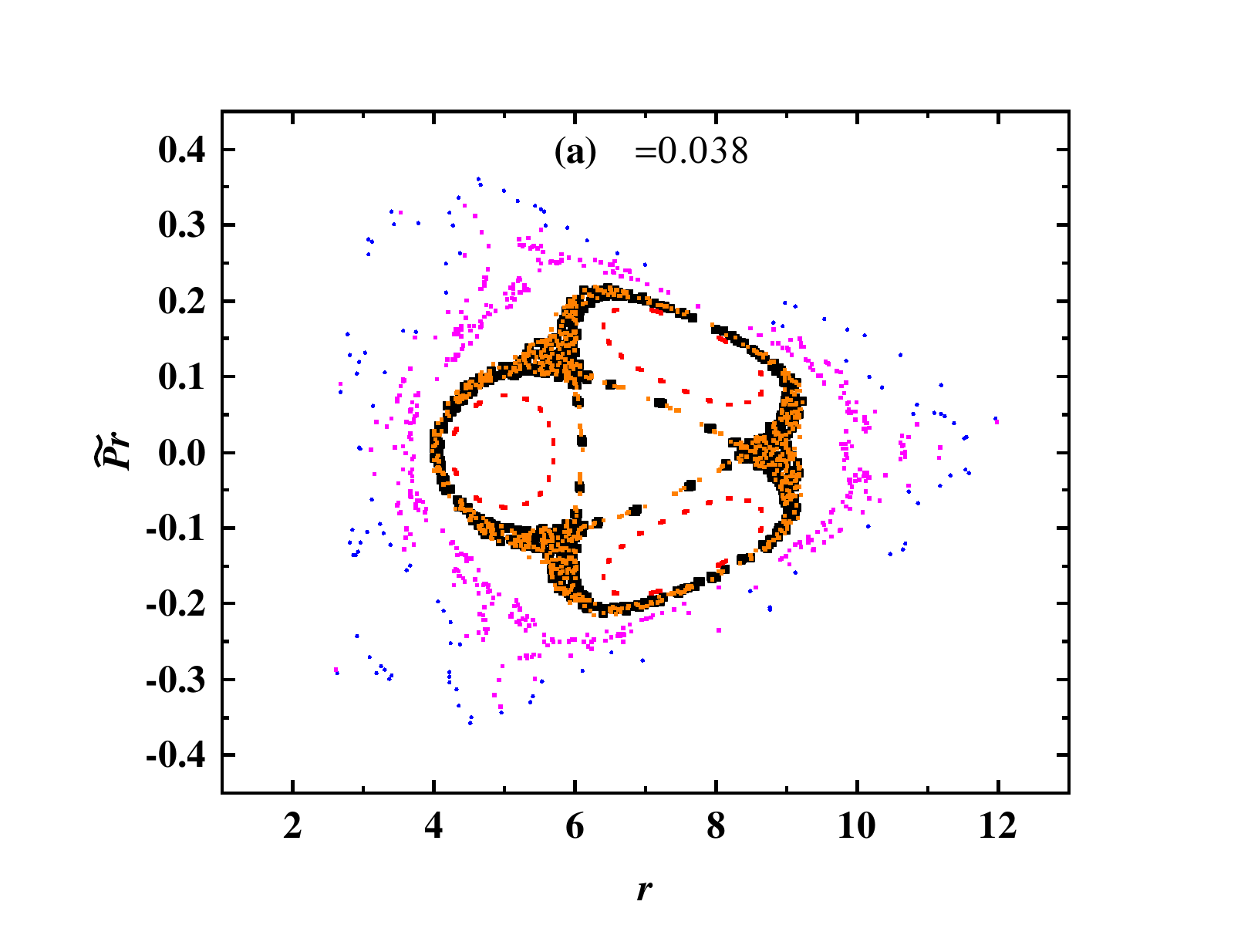}
    \includegraphics[width=20pc]{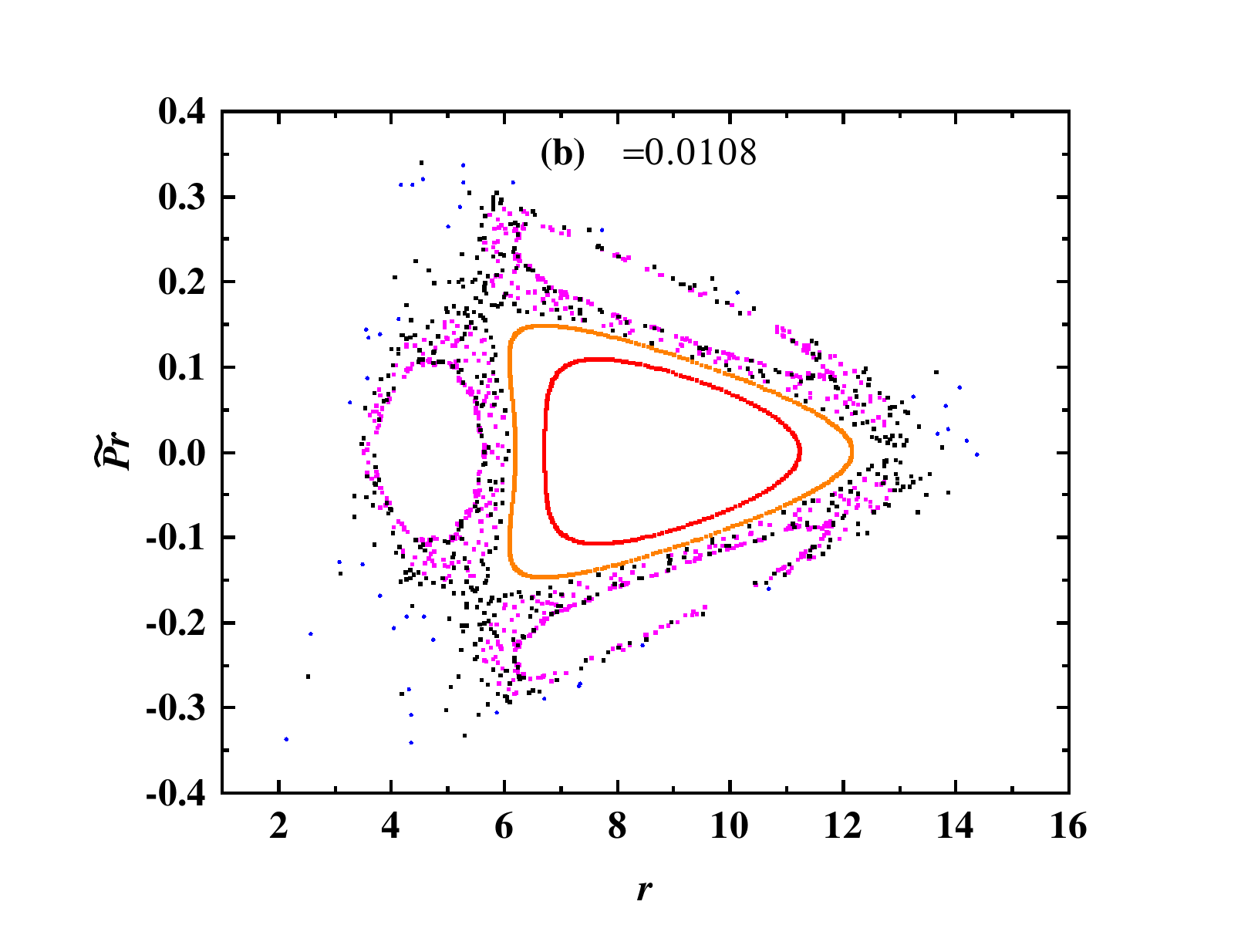}
    \includegraphics[width=20pc]{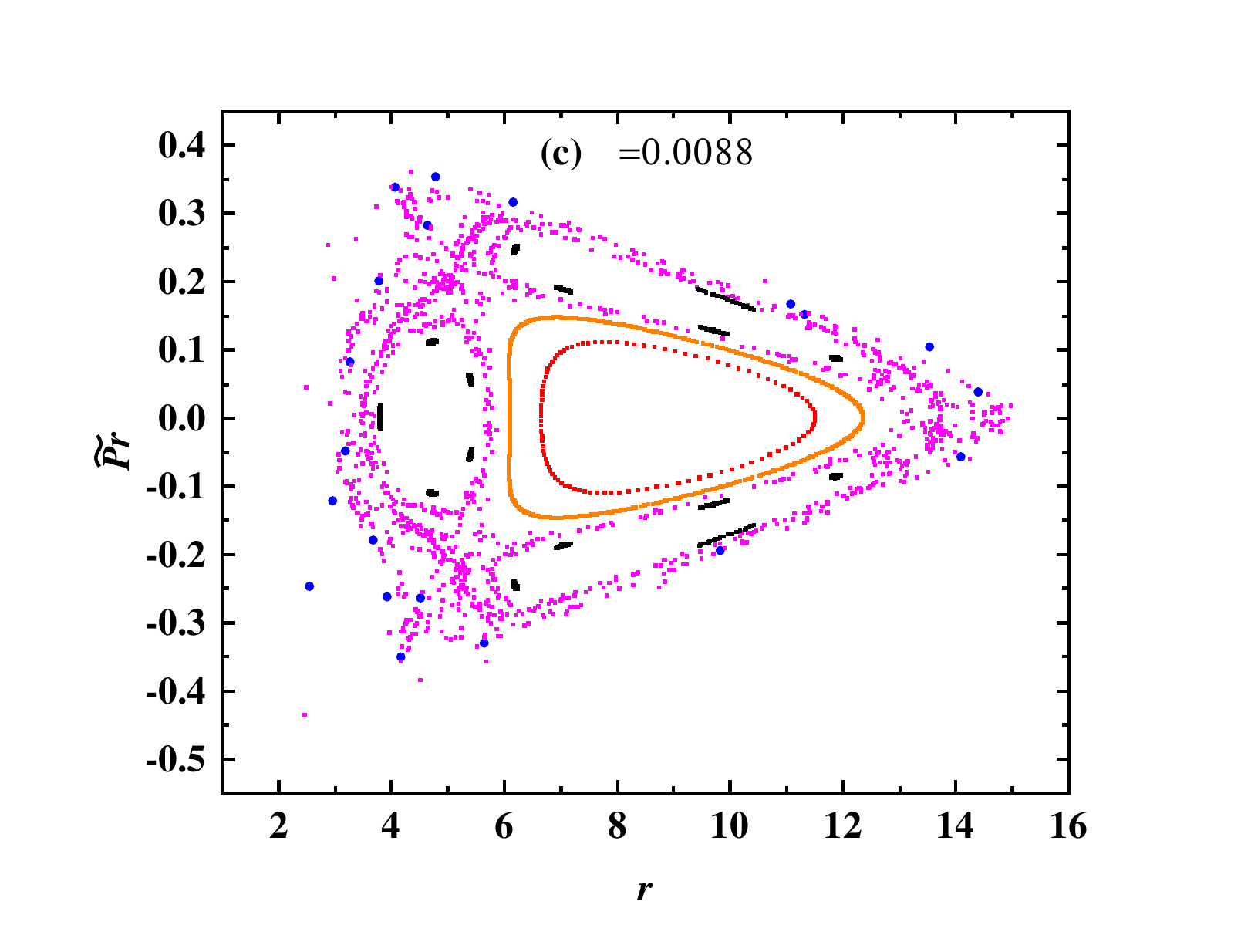}
    \includegraphics[width=20pc]{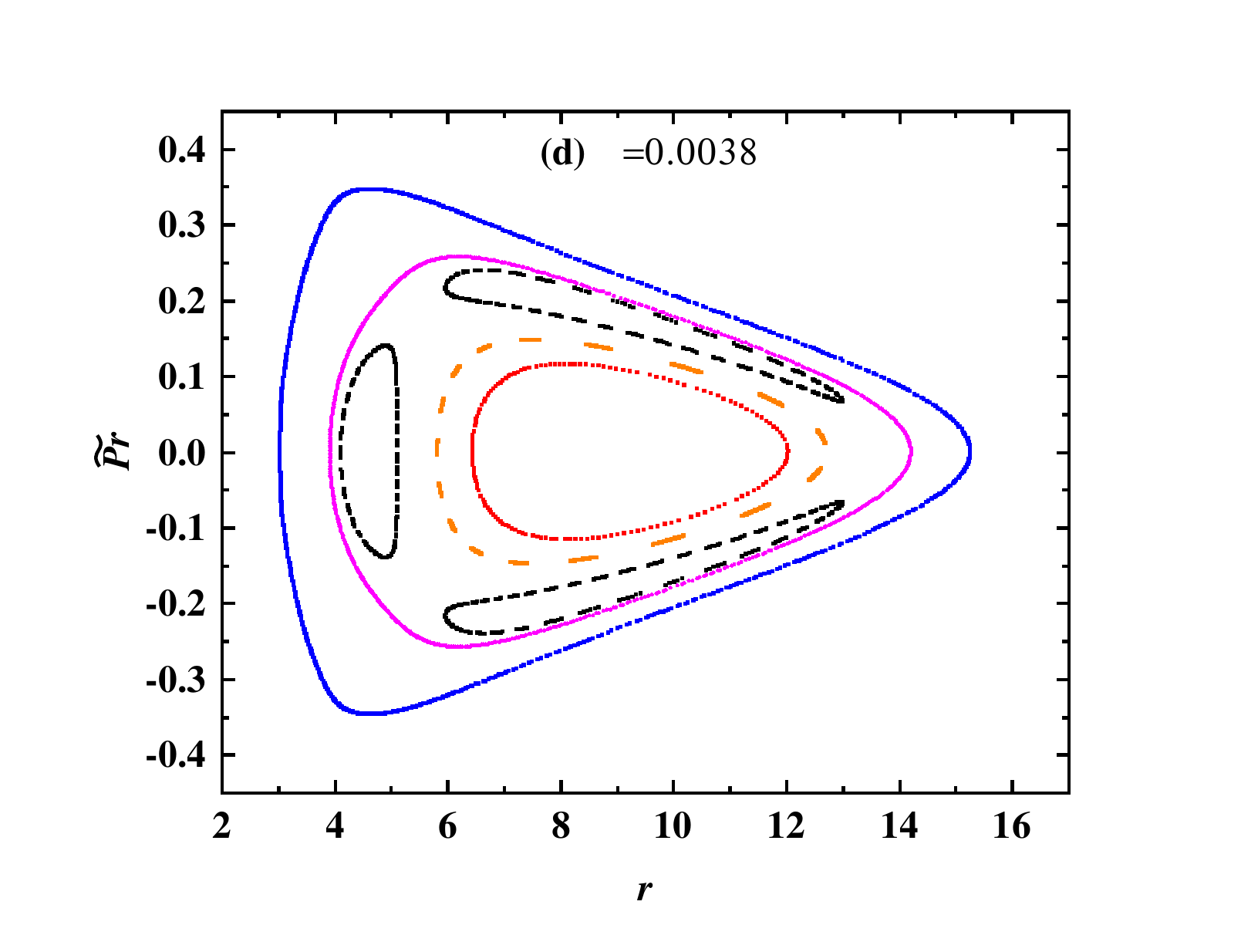}
\caption{Poincar\'{e} sections of massive particle orbits in the
magnetized KBR metric. The parameters are $a=0.8$,
$\tilde{E}=0.95$ and $\tilde{L}=2.5$. The magnetic field strength
$B$ is given four values.}}
\end{figure*}

\begin{figure*}[htpb]
    \centering{
\includegraphics[width=20pc]{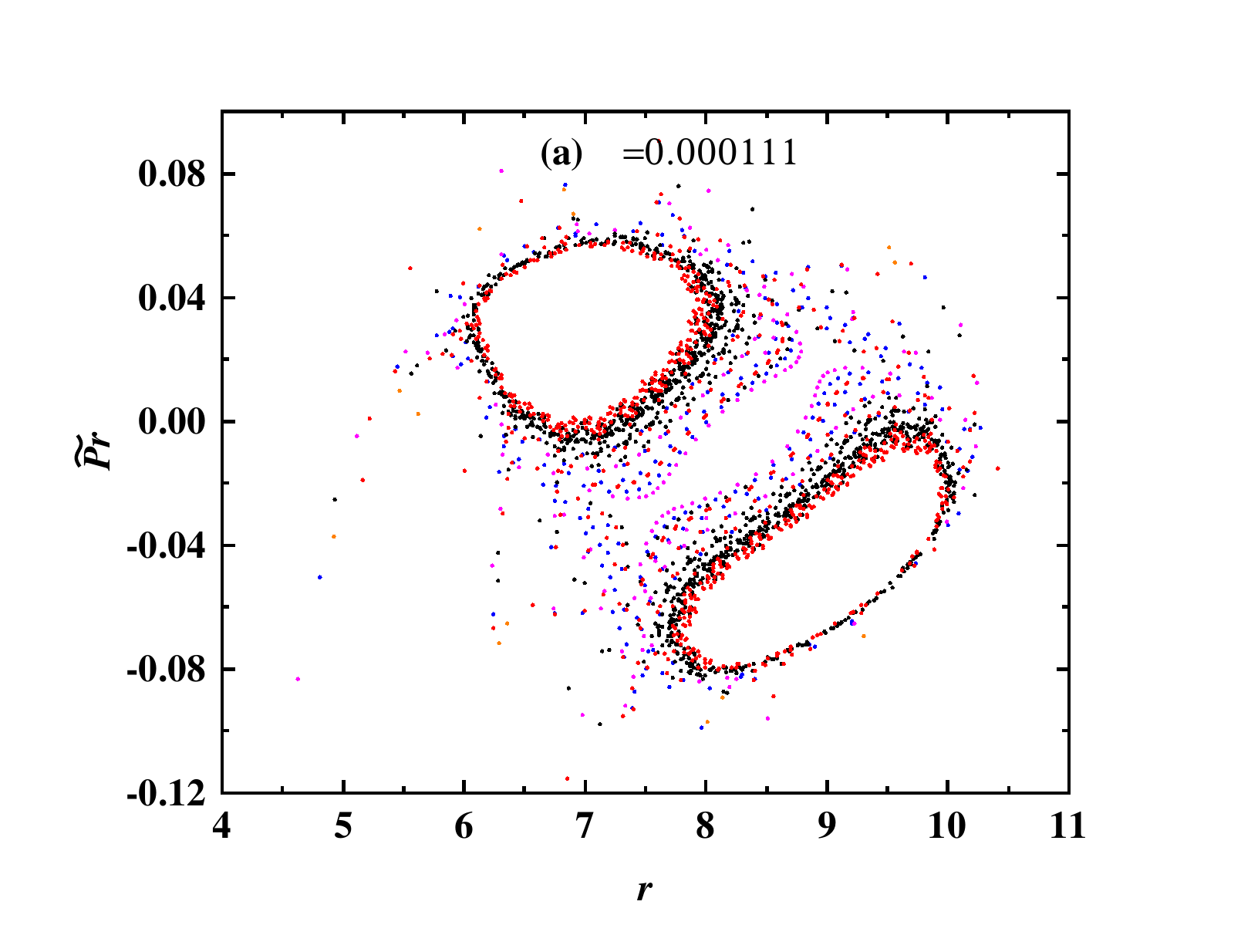}
\includegraphics[width=20pc]{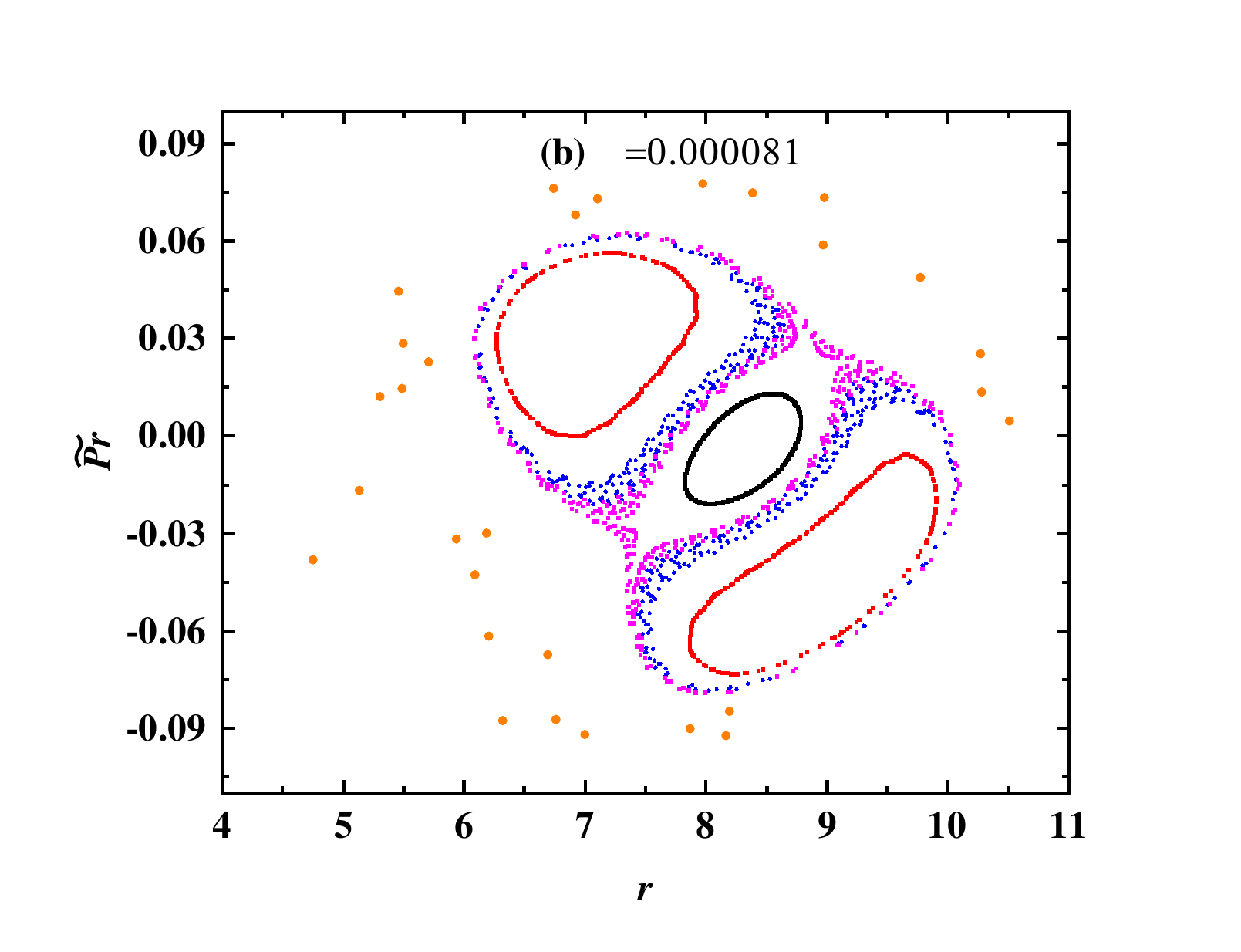}
\includegraphics[width=20pc]{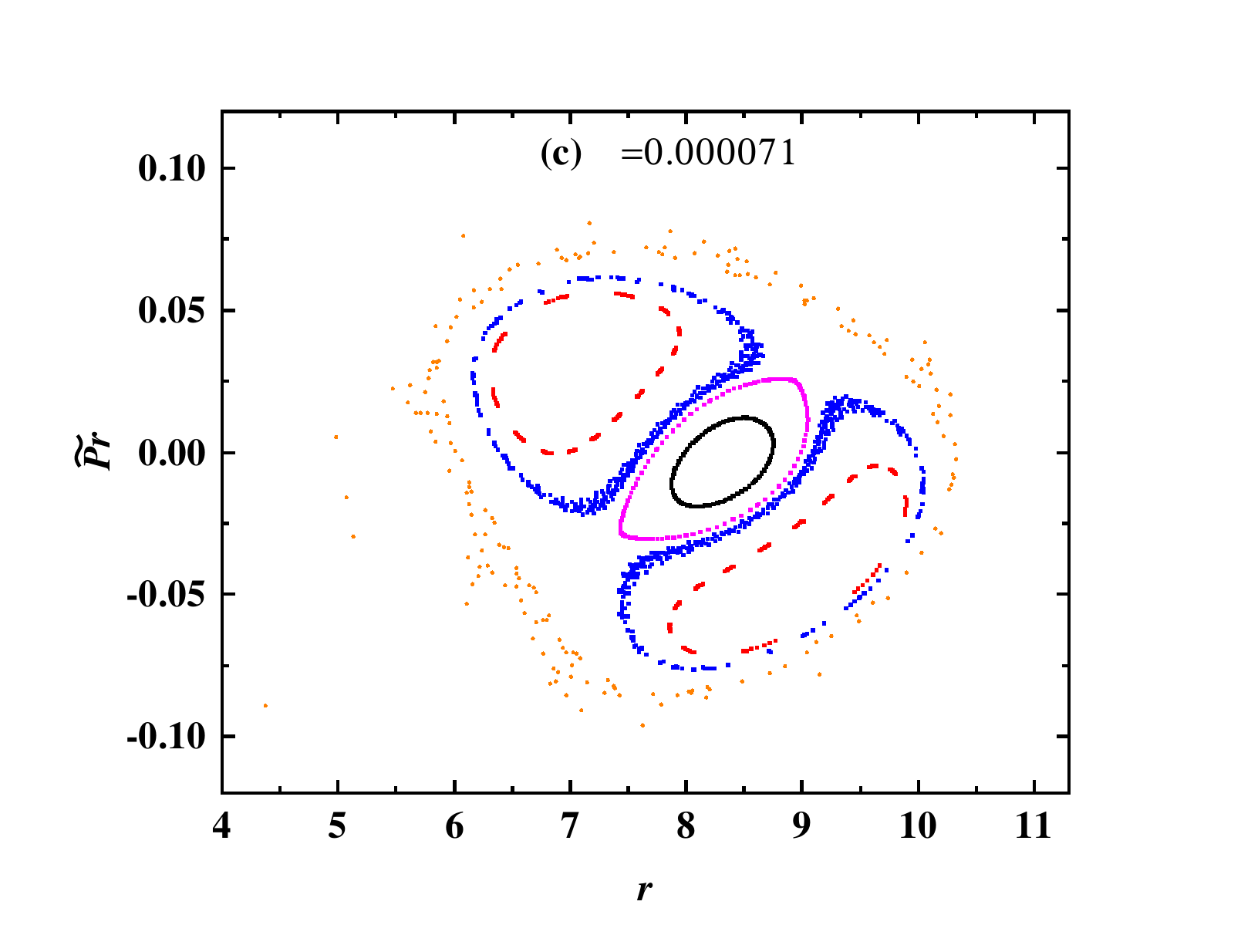}
\includegraphics[width=20pc]{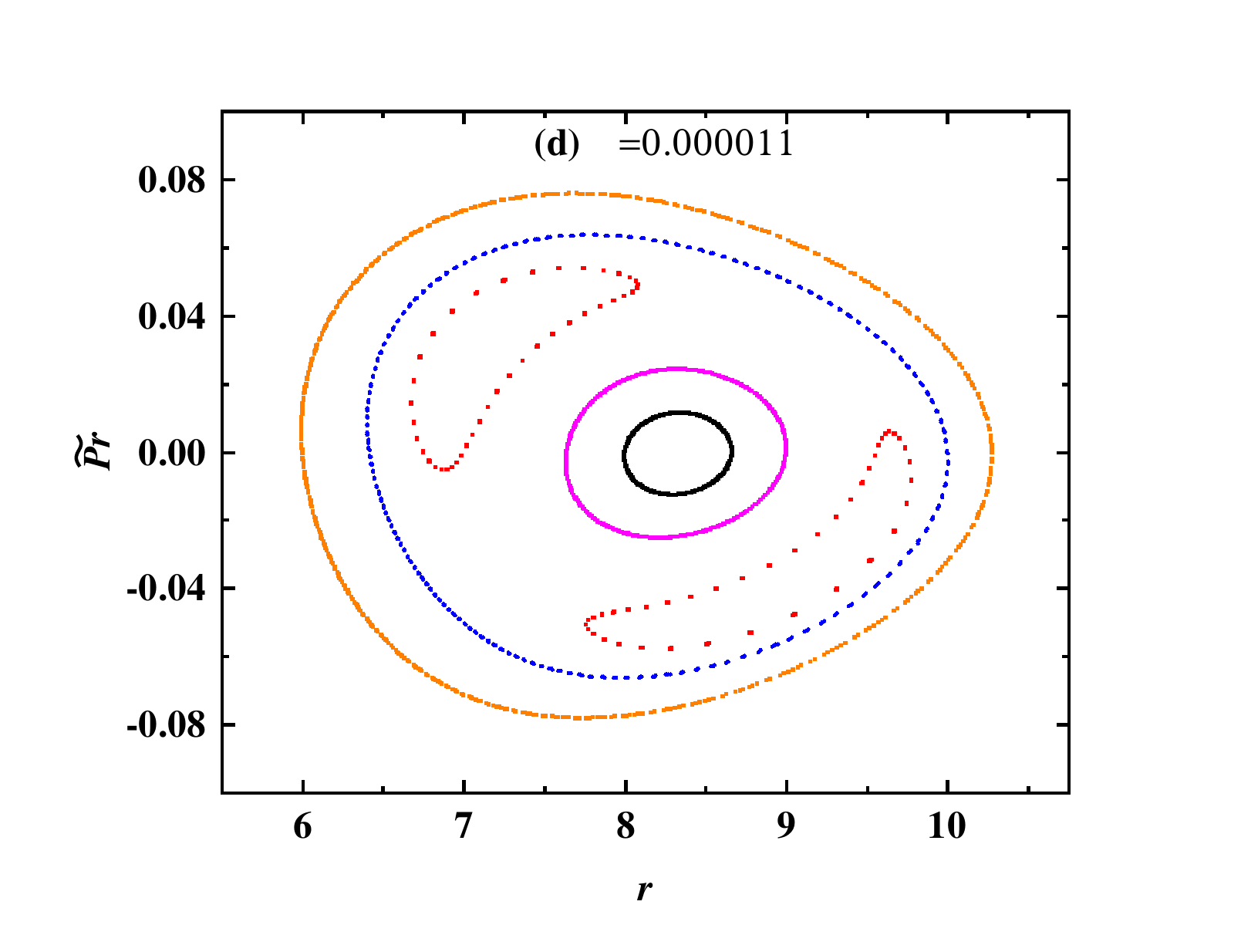}
\caption{Poincar\'{e} sections of massive particle orbits in the
accelerating Schwarzschild  metric. The parameters are
$\tilde{E}=0.95$ and $\tilde{L}=2.6$. The acceleration parameter
$A$ is given four values.}}
\end{figure*}

\end{document}